\documentclass[aps,prd,superscriptaddress,nofootinbib,floatfix,preprint]{revtex4}

\usepackage{graphicx}   

\newcommand{\beq}{\begin{equation}}
\newcommand{\eeq}{\end{equation}}
\newcommand{\beqa}{\begin{eqnarray}}
\newcommand{\eeqa}{\end{eqnarray}}
\newcommand{\bpr}{\begin{problem}}
\newcommand{\epr}{\end{problem}}
\newcommand{\bcent}{\begin{center}}
\newcommand{\ecent}{\end{center}}
\newcommand{\bfig}{\begin{figure}}
\newcommand{\efig}{\end{figure}}
\newcommand{\bpc}{\begin{picture}}
\newcommand{\epc}{\end{picture}}
\newcommand{\barr}{\begin{array}}
\newcommand{\earr}{\end{array}}
\newcommand{\bitm}{\begin{itemize}}
\newcommand{\eitm}{\end{itemize}}
\newcommand{\bright}{\begin{flushright}}
\newcommand{\eright}{\end{flushright}}
\newcommand{\bminip}{\begin{minipage}}
\newcommand{\eminip}{\end{minipage}}
\newcommand{\btab}{\begin{tabular}}
\newcommand{\etab}{\end{tabular}}

\newcommand{\nnb}{\nonumber}

\newcommand{\hiroshima}{Graduate School of Science, Hiroshima University, Kagamiyama, Higashi-Hiroshima 739-8526, Japan}

\newcommand{\izest}{International Center for Zetta-Exawatt Science and Technology, Ecole Polytechnique, Route de Saclay, Palaiseau, F-91128, France}

\begin{document}

\title{
The first search for sub-eV scalar fields via
four-wave mixing at a quasi-parallel laser collider
}
\author{Kensuke Homma} \affiliation{\hiroshima}\affiliation{\izest}
\author{Takashi Hasebe} \affiliation{\hiroshima}
\author{Kazuki Kume} \affiliation{\hiroshima}

\date{\today}

\begin{abstract}
A search for sub-eV scalar fields coupling to two photons has been performed 
via four-wave mixing at a quasi-parallel laser collider for the first time. 
The experiment demonstrates the novel approach
to search for resonantly produced sub-eV scalar fields
by combining two-color laser fields in the vacuum.
The aim of this paper is to provide the concrete
experimental setup and the analysis method based on specific combinations
of polarization states between incoming and outgoing photons,
which is extendable to higher intensity laser systems operated
at high repetition rates.
No significant signal of four-wave mixing was observed
by combining a $0.2\mu$J/0.75ns pulse laser and a 2mW CW laser on the
same optical axis.
Based on the prescription developed for this particular experimental approach,
we obtained the upper limit at a confidence level of 95\%
on the coupling-mass relation.
\end{abstract}

\maketitle

\section{Introduction}
A large fraction of dark components in the universe motivates us 
to search for yet undiscovered fields to naturally interpret the relevant 
observations. In high energy scales, several bosons have been discovered,
which can be understood as a result of spontaneous symmetry 
breaking such as pions 
based on chiral symmetry at the 0.1GeV scale and W/Z bosons based on gauge 
symmetry via the Higgs mechanism at the 100GeV scale. The relevant experimental
results show the evidence of coupling to two-photons of boson states,
for instance, via the decay process of the neutral pion and the Higgs-like particle.
These facts encourage further experimental searches for similar type of 
fields via two-photon coupling
in very different energy scales even apart from any theoretical 
speculations. In addition, there are actually theoretical rationales to 
expect sub-eV particles such as the axion (pseudoscalar boson)~\cite{Axion} 
and the dilaton (scalar boson)~\cite{STTL} associated with breaking of
fundamental symmetries in the context of particle physics and cosmology.
Therefore, we are led to probe such fields via their coupling to 
two-photons in the sub-eV mass range. 
Furthermore, the advent of high-intensity laser systems and the rapid
leap of the intensity encourage the approach
to probe weakly coupling dark fields with 
optical photons by the enhanced luminosity factor~\cite{ISMD2011,TajimaHomma}.

\begin{figure}
\bcent
\includegraphics[width=12.0cm]{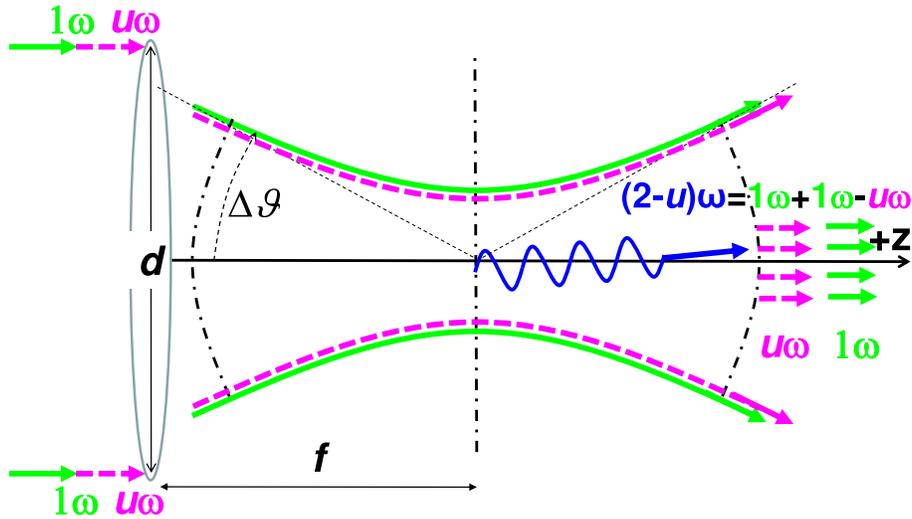}
\caption{
Quasi parallel colliding system (QPS) between two incident photons
out of a focused laser beam with the focal length $f$,
the beam diameter $d$, and the upper range of incident angles
$\Delta\vartheta$ determined by geometric optics.
The signature $(2-u)\omega$ is produced via
the four-wave mixing process,
$1\omega+1\omega\rightarrow(2-u)\omega+u\omega$ with $0<u<1$
by mixing two-color waves with different frequencies $1\omega$ and $u\omega$
in advance at the incidence.
}
\label{Fig1}
\ecent
\end{figure}

We advocated the concept of the quasi-parallel laser collider to 
produce a resonance state of a hypothetical boson in the sub-eV mass range
and simultaneously induce the decay in the background of a coherent laser field
~\cite{DEptp,DEapb,DEptep}. The exchange of such a low-mass field is interpreted 
as the four-wave mixing process in the vacuum.
Figure \ref{Fig1} illustrates the four-wave mixing process where
two photons with the degenerate energy $\omega$ are used for the
resonance production and the inducing laser field has the energy $u\omega$
with $0<u<1$, and a photon with the energy $(2-u)\omega$ is created 
as the signature of the interaction.
This four-wave mixing process is well-known in quantum optics~\cite{FWM,Yariv},
where atomic dynamics governs the phenomenon instead of the
exchange of the hypothetical resonance state, and
the application to test the higher-order QED effect can also be found
in Ref.~\cite{FWMqed}.
We identify this wave mixing system as a special kind of
photon-photon colliders intentionally, whose significant advantage 
is that the interaction rate has
cubic dependence on the laser intensity, while the conventional
particle colliders have quadratic dependence on the number of
charged particles per colliding bunch~\cite{DEptep}.
The cubic dependence of the interaction rate highly motivates us to extend
the method to much higher pulse energy and higher-intensity laser systems.

The proposed method can in principle distinguish whether the exchanged boson is
scalar or pseudoscalar based on combinations of
polarization states in the initial and final state photons \cite{DEptep}.
Therefore, the method has impacts on searches for
axion-like particles as well as the fifth force\cite{FifthForce}. 
In this paper, however, we focus only on the case for the scalar field exchange,
because the experimental setup is simpler.
The aim of this paper is to demonstrate the pilot experiment
to search for sub-eV scalar fields via the four-wave mixing process in the vacuum,
in which basic elements necessary for the proposed
experimental method and the data analysis are provided
so that they can be extended to much higher-intensity systems
operated at high repetition rates in the near future~\cite{ICAN}.

\section{Coupling-mass relation}
In this section we summarize formulae necessary to obtain the coupling-mass
relation from experimental parameters. The formulae are basically from Ref.
\cite{DEptep}, however, we re-evaluate the relations to apply to the realistic 
experimental conditions in the pilot search. We thus provide the details of 
the corrections in Appendix A and B compared with those in Ref.\cite{DEptp,DEptep}.

The effective interaction Lagrangian between two photons and
a hypothetical low-mass scalar field has the generic form
expressed as
\beqa\label{eq_phisigma}
-L_{\phi}=gM^{-1} \frac{1}{4}F_{\mu\nu}F^{\mu\nu} \phi
\eeqa
where $M$ has the dimension of energy while $g$ being a dimensionless constant.

\begin{figure}
\bcent
\includegraphics[width=10.0cm]{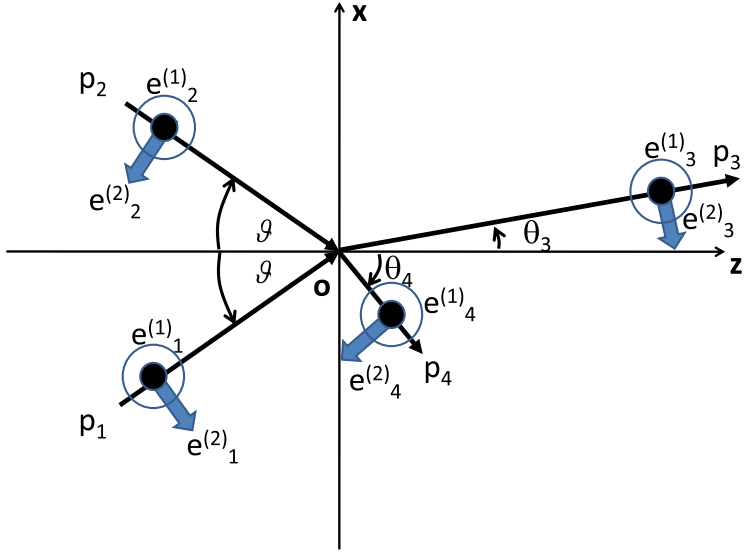}
\caption{
Definitions of kinematic variables~\cite{DEptp}.
}
\label{Fig2}
\ecent
\end{figure}

In the case of the scalar field exchange,
the possible linear polarization states in the four-wave mixing process
when all wave vectors are on the same plane as illustrated in Fig.\ref{Fig2}
are expressed as follows:
\begin{eqnarray}\label{eq_S}
\omega\{1\}+\omega\{1\} \rightarrow (2-u)\omega\{1\} + u\omega\{1\} \\ \nnb
\omega\{1\}+\omega\{1\} \rightarrow (2-u)\omega\{2\} + u\omega\{2\},
\end{eqnarray}
where
photon energies from the initial to the final state are denoted by
the linear polarization states \{1\} and \{2\} which are orthogonal to each other.
In this pilot experiment, we pursuit the second case of Eq.(\ref{eq_S}).

We then introduce notations to describe kinematics of four photons
as illustrated in Fig.\ref{Fig2}, where
the incident angle $\vartheta$ is assumed to be symmetric
around the $z$-axis in the average sense, 
because we focus lasers symmetrically by a lens element.
We introduce an arbitrary number $u$ with $0<u<1$
to re-define momenta of the final state photons as
\beq\label{eq000}
\omega_4 \equiv u\omega \quad \mbox{and} \quad \omega_3 \equiv (2-u)\omega,
\eeq
where we require $0 < \omega_4 < \omega_3 < 2\omega$.
We consider to measure $\omega_3$ with the specific
polarization states as the signature of the interaction.
With these definitions, energy-momentum conservation
is expressed as~\cite{DEptep}
\beqa\label{eq001}
(2-u)\omega + u\omega = 2\omega,
\eeqa
\beqa\label{eq002}
(2-u)\omega\cos\theta_3 + u\omega\cos\theta_4 = 2\omega\cos\vartheta,
\eeqa
\beqa\label{eq003}
(2-u)\omega\sin\theta_3 = u\omega\sin\theta_4,
\eeqa
\beqa\label{eq004}
{\mathcal R} \equiv \frac{\sin\theta_3}{\sin\theta_4} =
\frac{\omega_4}{\omega_3} = \frac{u}{2-u} =
\frac{\sin^2\vartheta}{1-2\cos\vartheta\cos\theta_4+\cos^2\vartheta}.
\eeqa

Given a set of physical parameters of the scalar field exchange: 
mass $m$, coupling to two photons $g/M$, and the polarization dependent factor
${\mathcal F}_S$ with a specified set of linear polarizations $S$ as explained below,
the yield is parameterized as
\beq\label{eq22}
{\mathcal Y} \equiv
\frac{1}{64\sqrt{2}\pi^4}
\left(\frac{\lambda_c}{c\tau_c}\right)
\left(\frac{\tau_c}{\tau_i}\right)
\left(\frac{f}{d}\right)^3
\tan^{-1}\left(\frac{\pi d^2}{4\ f\lambda_c}\right)
\frac{(\overline{u}-\underline{u})^2}{\overline{u}\underline{u}}
\left(\frac{gm{\mbox[eV]}}{M{\mbox[eV]}}\right)^2 \left(\frac{m{\mbox[eV]}}{\omega{\mbox[eV]}}\right)^3
{\mathcal W}{\mathcal F_S} C_{mb} N^2_c N_i,
\eeq
which is quoted from Eq.(\ref{eqY}) in Appendix A,
where 
the subscripts $c$ and $i$ denote creation and inducing lasers, respectively,
$\lambda$ wavelength, $\tau$ pulse duration, $f$ a common focal length,
$d$ a common beam diameter, $\overline{u}$ and $\underline{u}$
upper and lower values on $u$ determined by the spectrum width
of $\omega_4$, respectively,
$C_{mb}=1/2$ is the combinatorial factor originating from the consideration
on multimode frequency states~\cite{DEptep},
$N$ the average numbers of photons in coherent states,
${\mathcal W} \sim \pi/2$ the numerical factor relevant to the integral
of the weighted resonance function defined by Eq.(\ref{eq_calW}) in Appendix A,
and we partially apply natural units to parameters specified with [eV].

We have discussed the effect of the spectrum width
in the case that the spectrum width of the creation laser is 
negligibly small compared to that of the inducing laser in Ref.\cite{DEptep}.
However, this is not always the case. For example, this pilot experiment
is actually performed with the condition that both widths are nearly equal
as we explain in the next section.
We provide the treatment on the effective choice of $\overline{u}$ 
and $\underline{u}$ applicable to the most general case
in Appendix B in detail.

In the case of $S=1122$ as specified by the second of Eq.(\ref{eq_S}) 
applied to the scalar field exchange ($SC$), 
the polarization dependent factor is estimated as \cite{DEptep}
\beq\label{eq_Fs}
{\mathcal F}^{SC}_{1122} \sim 2\pi\left(\frac{3}{8}+3\hat{{\mathcal R}^2}-
\hat{{\mathcal R}}\right)
\eeq
with $\hat{{\mathcal R}} \equiv \frac{1}{2}({\mathcal R}+{\mathcal R}^{-1})$
for a given $u$ via ${\mathcal R}$ in Eq.(\ref{eq004}).
This factor originates from a degree of freedom on the azimuthal rotation
of the plane including $p_3$ and $p_4$ around the $z$-axis 
in Fig.\ref{Fig2} with respect to that defined by $p_1$ and $p_2$.
The other polarization combinations can also be calculated based on Appendix
of Ref.\cite{DEptep}.

From Eq.(\ref{eq22}) we express the coupling
parameter $g/M$ to discuss the sensitivity as a function of $m$
for a given set of experimental parameters via the following equation
\beqa\label{eq24}
\frac{g}{M{\mbox[eV]}} = 
2^{1/4}8\pi^2 
\sqrt{
\frac{{\mathcal Y} \omega^3{\mbox[eV]} }
{
\left(\frac{\lambda_c}{c\tau_c}\right)
\left(\frac{\tau_c}{\tau_i}\right)
\left(\frac{f}{d}\right)^3
\tan^{-1}\left(\frac{\pi d^2}{4\ f\lambda_c}\right)
\frac{(\overline{u}-\underline{u})^2}{\overline{u}\underline{u}}
{\mathcal W}{\mathcal F_S} C_{mb} N^2_c N_i
}
}
m^{-5/2}{\mbox[eV]} 
\eeqa
under the condition $\tau_c \le \tau_i$, because the pulse energy
of the creation laser is more important to enhance the signal yield
due to the quadratic dependence on the energy.

\section{Experimental setup to detect four-wave mixing}

\begin{figure}
\bcent
\includegraphics[width=16.0cm]{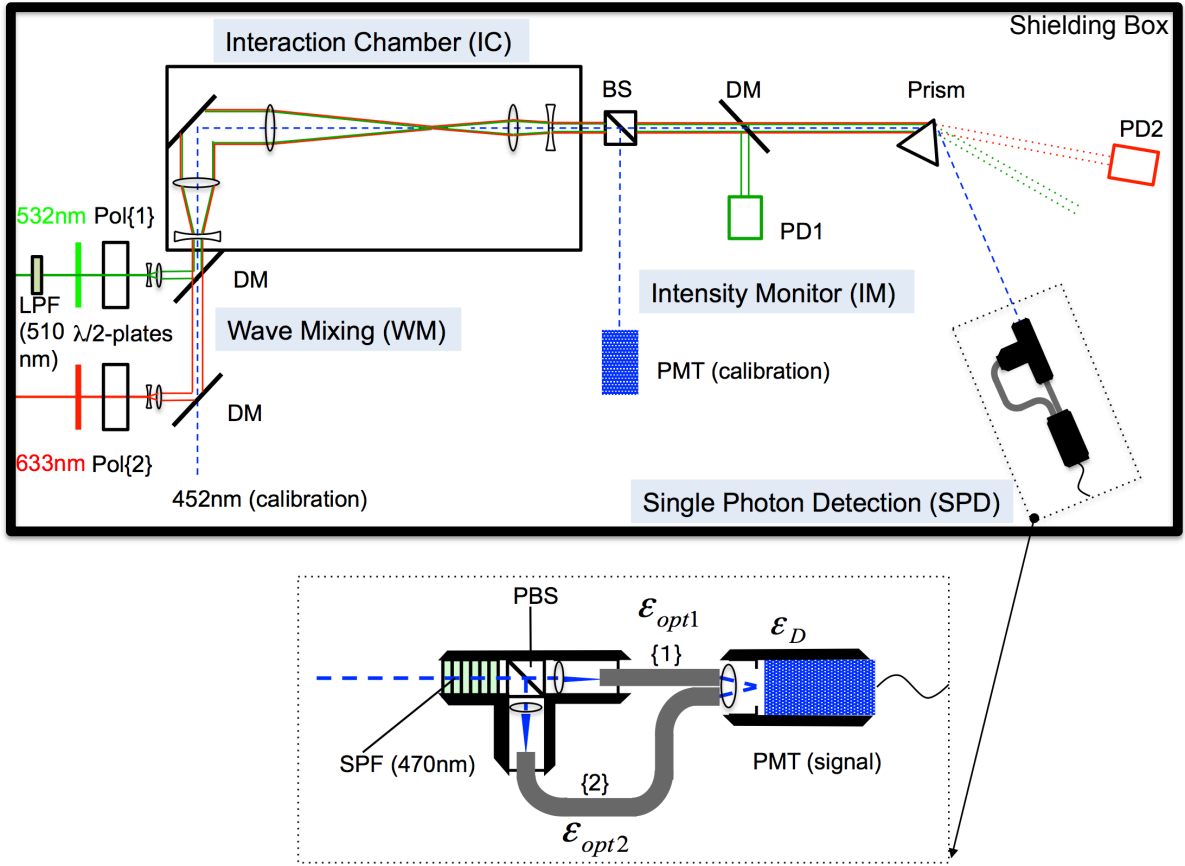}
\caption{
Schematic view of the experimental setup to measure
the four-wave mixing process in the vacuum.
}
\label{Fig3}
\ecent
\end{figure}

Figure \ref{Fig3} shows the schematic view of the experimental setup.
The setup consists of the wave mixing part (WM), the interaction vacuum
chamber (IC), the intensity monitoring part (IM) and the single photon 
detection part (SPD).
 
WM combines a pulsed laser beam to create a resonance state
with a continuous wave (CW) laser beam which induces the decay
with specified linear polarization states.
For the purpose of the calibration, a laser beam
containing the signal energy $(2-u)\omega$ is also combined.
The wave mixing can be achieved by a set of dichroic mirrors.
  
The pulsed laser is the linearly polarized NanoLaser, 
a monolithic passively Q-switched microchip laser
.
The laser cavity consists of a $2$mm${}^3$ Nd:YAG gain medium bonded to 
a chromium doped YAG saturable absorber.
The repetition rate depends on the diode pumping power:
the higher the pumping power, the faster the absorber saturates
at which the absorber becomes transparent. 
The measured repetition rate is 18.5~kHz.
The cavity's mirrors are vapor deposited on the both sides of the
crystal to form a monolithic oscillator pumped by a CW diode laser.
The second harmonic wavelength 532nm is produced by converting
the fundamental wavelength 1064nm.
The pulse energy of the 532nm wave is typically $0.8\mu$J per 0.75 ns 
pulse duration with the transverse mode close to TEM00 at the ejection 
point of the laser. 
  
The CW laser is the linearly polarized Helium-Neon laser at the wavelength 
632.8 nm with the power 10mW at the ejection point of the laser
and the transverse mode TEM00.
  
Relative line widths of these lasers with respect to their central
frequencies are summarized in Tab.\ref{Tab2}, respectively.
In addition, as we discuss in Appendix B about the relation on line widths 
between creation and inducing lasers which is relevant to
$\overline{u}$ and $\underline{u}$ in Eq.(\ref{eq22}),
we evaluate the effective line width in the averaged quasi-parallel frame
based on Eq.(\ref{eq_convou}) in Appendix B. 

Given wavelengths for creation laser ($\lambda_c = 532$ nm)
and inducing laser ($\lambda_i = 633$ nm) beams which result in
$u=\lambda_c/\lambda_i=0.84$,
the corresponding wavelength of four-wave mixing is expected to be
\beq\label{eq_wavelength}
\lambda_s = \frac{\lambda_c/2 \cdot \lambda_{i}}{\lambda_i - \lambda_c/2}
= 459 \mbox{ nm}.
\eeq
Thus, in addition to the creation and inducing laser beams, a blue diode laser
covering the $452 \pm 10$nm wavelength range is used to supply the calibration
source to align the detection system and to obtain the detector efficiency
with respect to the signal photon of $(2-u)\omega$. 

The linear polarizers made of calcite with the extinction ratio of $O(10^{-4})$
specifies the incident linear polarization states of the creation
and inducing laser fields as well as the calibration laser field, respectively.
The linear polarization state $\{1\}$ is determined by the polarizer (Pol\{1\})
for the creation laser, while the state $\{2\}$ is set by the polarizer
(Pol\{2\}) for the inducing laser 
which is adjusted so that the planes of the linear
polarizations become orthogonal to each other. In front of the linear polarizers,
$\lambda/2$-plates are placed and adjusted so that the transmittance of 
almost linearly polarized lasers at the output of the laser systems is 
maximized. In front of the ejection point of the creation pulse laser,
we put a long pass filter to accept wavelength only above 510~nm to
suppress wavelength close to the signal wavelength 459~nm in advance of
the wave mixing with the optical density OD$\sim 4$ where OD is defined as
OD$\equiv-\log_{10}(I/I_0)$ with output and input photon intensities
$I$ and $I_0$, respectively.

After wave mixing with specified polarization states, the combined laser
beams share the common optical axis, and they are guided into the interaction
chamber (IC) maintained at $1.2 \times10^{-4}$Pa. 
Inside IC, a set of achromatic lenses expands the beam diameter 
to 40 mm and focuses the combined beams with the focal length $f=200$ mm.
The focal spots are monitored by a beam profile monitor and the centers
of the two-color beam profiles at a point near from the common focal spot
are aligned to each other with $\sim 10\mu$m precision.
The agreement of the optical axes between the two beams are also confirmed
by checking the profile overlap at a different point from the focal spot
along the common optical axis.
After focusing the combined beams inside IC, 
the divergent beams are parallelized by a set of achromatic lenses
with the reduced beam diameter 10~mm and guided to outside IC
through the chamber window. 
The beam focusing parameter is important, because this gives the upper limit
of the sensitive mass range by the possible range of the incident angle 
$\vartheta$ via Eqs.(\ref{eq15}) - (\ref{eq_rho}) in Appendix A.
The resonance production is enhanced when $m = 2\omega\sin\vartheta$ is
satisfied, that is, when a CMS-energy between two incident photons
coincides with the exchanged mass. 
If an angular coverage $\Delta\vartheta$ is too small compared to
a mass we are interested in, the resonance condition is never satisfied.
Therefore, the focusing parameter can adjust
the sensitive mass range via the upper limit on the incident angles. 
In this pilot experiment the upper limit on mass
is thus 0.46~eV for the creation laser wavelength of 532~nm (2.3~eV).
%

The absolute beam intensities are monitored by the beam profile monitor.
The non-interacting
creation and inducing laser fields are kicked out by individual
dichroic mirrors and the reflected waves are measured by the photo diodes
PD1 and PD2 for 532 nm and 633 nm, respectively.
These amplitude information are used to monitor the relative intensity
variation on the shot-by-shot basis.

The signal wave is further guided to the equilateral prism made of N-SF11 glass
with the MgF${}_2$ anti-reflection coat
and refracted to the detection system, while the residual non-interacting
creation and inducing beams are refracted to different directions.
These non-interacting waves are further reflected by mirrors and dumped 
apart from the detection point of the signal wave.

\begin{figure}
\bcent
\includegraphics[width=12.0cm]{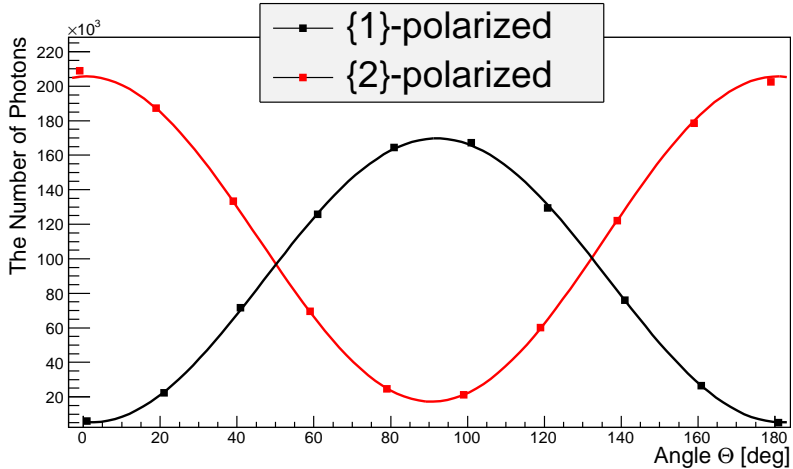}
\caption{
The number of photons in $T\{1\}$ and $T\{2\}$
as a function of the relative rotation angle
between the polarization plane
of the pulsed creation laser and that of the 
polarization beam splitter (PBS).
}
\label{Fig3pol}
\ecent
\end{figure}

For the scalar field search, we require the initial and final state 
photon energies with their linear polarization states as follows
\beq\label{eq_true}
\omega\{1\} + \omega\{1\} \rightarrow u\omega\{2\} + (2-u)\omega\{2\},
\eeq
where
$\{1\}$ and $\{2\}$ specifies the orthogonal linear polarization states
for photons in the creation beam and in the inducing beam, respectively.
As a reference polarization combination, we also measure the following case simultaneously
\beq\label{eq_false}
\omega\{1\} + \omega\{1\} \rightarrow u\omega\{2\} + (2-u)\omega\{1\},
\eeq
which is not allowed when all wave vectors in Fig.\ref{Fig2} are on the same plane.
Due to a degree of freedom on the relative rotation angle between $p_1-p_2$ and 
$p_3-p_4$ planes around the $z$-axis, however,
there is a finite probability to accept this polarization combination.
In order to measure the both cases on the shot-by-shot basis, 
we introduced the polarization beam splitter (PBS) 
whose polarization directions are aligned to the incident laser 
polarizations $\{1\}$ and $\{2\}$ in advance, respectively
(see the enlarged view of the SPD part in Fig.\ref{Fig3}).
For the two polarization paths behind the PBS, two plastic optical fibers
whose transmittance are independent of the incident linear polarization states
are attached with different lengths by introducing a relative time delay 23.75~ns.
These two optical fibers are attached to a common photomultiplier tube (PMT), 
which is a metal package PMT with the rise time 0.78~ns
(R7400-01 manufactured by HAMAMATSU),
through lenses so that we can count the number
of photons in the two different time domains $T\{1\}$ and $T\{2\}$
separated by that time difference on the digitized wave 
form of the analog output from the common PMT.
By using the fiber-coupled PMT,
the degree of the linear polarization is measured by rotating the polarization plane
of the pulsed creation laser with respect to that of the PBS.
Figure \ref{Fig3pol} shows how the number of photons in $T\{1\}$ and $T\{2\}$
change as a function of the relative rotation angle $\Theta$ in units of
degree, respectively.
The fit results with the functional form $\sin(\pi\Theta/180)$
are consistent with a nearly linear polarized state
after transmission though all of optics including dichroic mirrors.
%

In order to shutout residual non-interacting creation and inducing laser
photons, five short pass filters (SPF) with the nominal OD$\sim 4$ for each 
to accept only wavelength below 470~nm are placed inside the tube 
in front of the PBS.
The detection efficiency to the signal wavelength is evaluated 
in advance of the pilot measurement with the two same-type PMT's with 
the beam splitter(BS) in Fig.\ref{Fig3}.

The readout of the analog signal from the photomultiplier is performed
by 4-ch waveform digitizer (10-bit cPCI High-Speed Digitizers, Acqiris DC282 
type U1065A provided by Agilent Technologies)
without any electronics for amplification
to avoid adding noise sources. The measured maximum rate of the readout 
by requiring simple online preselections is $\sim10$~kHz. 
The digitizer is similar to the digital oscilloscope, however, the readout
speed is three orders of magnitude higher. Thus online selections based
on an algorithm are applicable before waveforms are actually stored.

Trigger signals are created by discriminating pulse heights
of analog signals on PD1 and the digitizer is synchronized with these trigger 
signals. By denoting the existence or absence of green (532 nm) 
and red (633nm) lasers
as $g$ and $r$ or $\bar{g}$ and $\bar{r}$, respectively, 
we can consider following four wave-mixing patterns:
$g+r$, $\bar{g}+\bar{r}$, $g+\bar{r}$, and $\bar{g}+r$ representing cases
including signal (S), dark currents or pedestal (P), 
residual of creation laser photons (C), 
and residual of inducing laser photons (I), 
respectively. The pedestal trigger is produced immediately
after every 18.5~kHz green triggers by adding a constant delay at which 
the green laser pulse is physically absent. 

We put a physical shutter on the CW red laser beam line
repeating open and close every 2 sec. 
Monitoring the green and red laser amplitudes at individual triggers
allows to identify the four patterns of wave-mixing
S, P, C and I based on the offline analysis on the recorded 
digitized waveform. Waveforms are recorded with 500 sampling 
points during the 250 ns time window corresponding to 0.5 ns/division
which is consistent with the time resolution on the leading edge of
the used photomultiplier(PMT) for the single photon detection.

As the online level trigger, we required that at least 
one signal-like signature below -1.00 mV threshold 
after online pedestal subtraction is found in
either $T\{1\}$ or $T\{2\}$ time domain
within the two 15 ns windows (see green bands in Fig.\ref{Fig4}
and \ref{Fig6}), and only waveforms containing such a 
signature are recorded on the disk for the offline analysis.

The number of total triggers reached $2.5 \times 10^9$ during the
pilot measurement over four days.

\section{Offline data analysis}
In order to test statistical significance of four-wave mixing signals,
the quantities we discuss are $N_{Si}$ 
which are acceptance-uncorrected numbers of photons found 
in the time domains $T\{i\}$ with the linear polarization states $i=1,2$ 
in the case of the signal pattern (S).
For the following paragraphs, we abbreviate
the symbols of the time domains with specified polarization states,
unless confusion is expected.

First, the four patterns: 
signal (S), pedestal (P), residual of green laser photons (C) and residual of 
red laser photons (I) are identified by looking at amplitudes of photodiodes 
recorded in the waveform data, and the number of events of individual patterns:
$W_S$, $W_P$, $W_C$ and $W_I$, respectively, are counted. 
These numbers are used as the weights to evaluate the number of
photons in the signal pattern by subtracting those in the other patterns.

Since there is no complete wave filters, we must expect non-zero numbers of
residual photons in the three patterns except the pedestal pattern 
where the dominant background is the thermal noise from the photomultiplier. 
We, therefore, interpret the observed raw numbers of photons $n$ 
in the four patterns specified with individual subscripts as
\beqa
n_S &=& N_P + N_C + N_I + N_S\nnb\\
n_C &=& N_P + N_C\nnb\\
n_I &=& N_P + N_I\nnb\\
n_P &=& N_P,
\eeqa
where we assume that the observed pedestal counts include thermal noises
from the photodevice and ambient noises such as cosmic rays, hence,
the pedestal counts should be commonly included in the other three patterns 
in the average sense.
By considering the event weights of the four trigger patterns,
we then deduce the true number of photons in the signal pattern as follows
\beqa\label{eq_signal}
N_S &=& 
n_S-\frac{W_S}{W_P}n_P-\frac{W_S}{W_C}(n_C-n_P)-\frac{W_S}{W_I}(n_I-n_P)\nnb\\
    &=& n_S-\frac{W_S}{W_C}n_C-\frac{W_S}{W_I}n_I+\frac{W_S}{W_P}n_P.
\eeqa
We note that, exactly speaking, the physical meaning of $n_P$ is different 
from the number of photons, because the dominant pedestal charges are produced
by thermal noises of the photodevice. As long as observed charges are
expressed in units of single-photon equivalent charge, however, this treatment
is justified.

\begin{figure}
\bcent
\includegraphics[width=14.0cm]{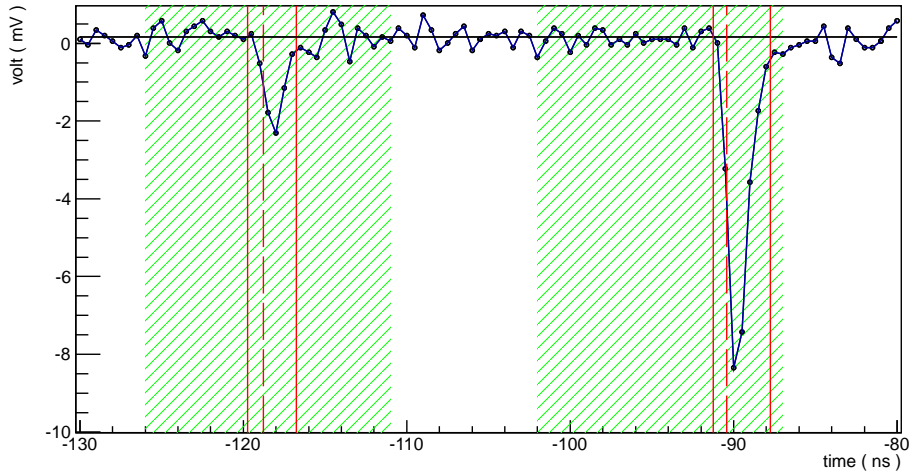}
\caption{
Single shot example of a digitized waveform with peaks
within the 50 ns time window.
}
\label{Fig4}
\ecent
\end{figure}

The analysis steps to obtain observed raw numbers $n_{ij}$ 
with trigger patterns $i=S,C,I,P$ and 
linear polarization states $j=1,2$ are as follows.
Figure \ref{Fig4} shows a single shot example of the digitized waveform
within a 50 ns time window.
The two time domains within a 3.5~ns interval 
subtended by two solid vertical lines, respectively,
are equally defined which are separated by
the known time difference of 23.75 ns due to the different optical
fiber lengths.
The shorter (earlier time domain) and longer (later time domain) fibers 
correspond to the linear polarization states $\{1\}$ and $\{2\}$, respectively.

Photon-like signals or thermal noise signals are identified by the
negative peak finding. After finding a time bin with the largest amplitude
in the negative direction, a global pedestal amplitude is determined
by averaging over the 250~ns window excluding the peak region
as shown by the horizontal line in Fig.\ref{Fig4}.
We then find the falling edge and define the signal arrival time $t_0$ 
at the detector as the time bin at 
the half value of the peak amplitude after subtracting the pedestal value, 
which is indicated by the dotted vertical line.
We require that a peak structure is identified by a pair of falling
and rising edges around the peak position $t_p$. By defining time 
intervals from the falling edge to the peak and the peak to the rising edge
as $\Delta t_f$ and $\Delta t_r$, respectively,
the time window of a signal, $t_{sig}$, is defined as 
$t_p - 2.0\Delta t_f \le t_{sig} \le  t_p + 2.25\Delta t_r$
which are indicated by solid vertical lines.
The charge sum in that time window is evaluated 
in units of single-photon equivalent charge $-4.21\times 10^{-14}$~C.

The single-photon equivalent charge from the used PMT operated 
at -800~V was evaluated with weak pulsed photon sources ranging from 0 to 
several tens photons as the average number of incident photons 
per pulse injection.
Based on the Poissonian probability distribution for a given set
of the pulsed photon sources, we can estimate the expected charge 
when a single photon is injected so that this charge is common to
all the weak photon sources.

\begin{figure}
\bcent
\includegraphics[width=14.0cm]{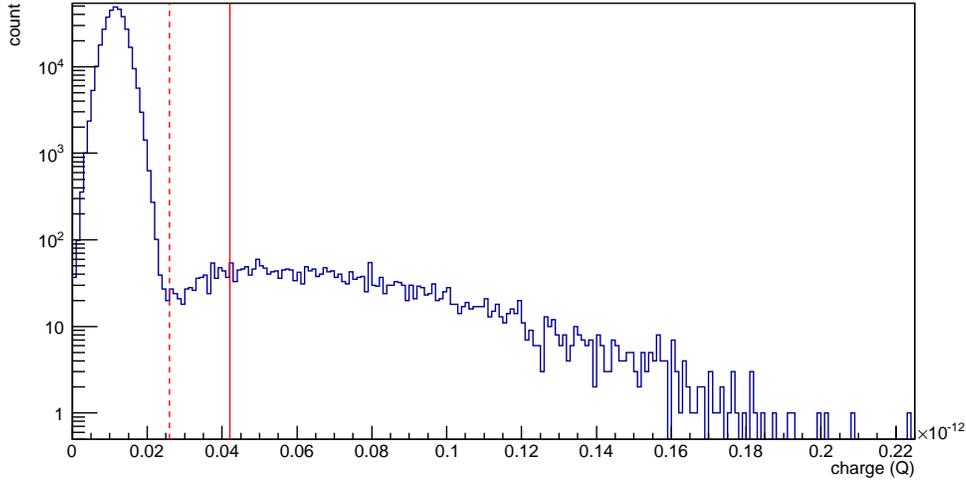}
\caption{
Charge sums within time windows of photon-like signals 
in the all four trigger patterns.
The offline cut is indicated by the dotted vertical line
with respect to the solid vertical line corresponding to
single-photon equivalent charge.
}
\label{Fig5}
\ecent
\end{figure}

We then choose the waveforms satisfying the condition that at least 
one peak above 0.6 photon equivalent charge
($-2.60\times 10^{-14}/-4.21\times 10^{-14}$)
is contained for counting the number of photons in the four patterns.
The offline cut is indicated by the dotted vertical line in Fig.\ref{Fig5}
with respect to single-photon equivalent charge denoting by 
the solid vertical line,
where the charge sums within time windows of photon-like signals
are shown in the all four trigger patterns.

\begin{figure}
\bcent
\includegraphics[width=16.0cm]{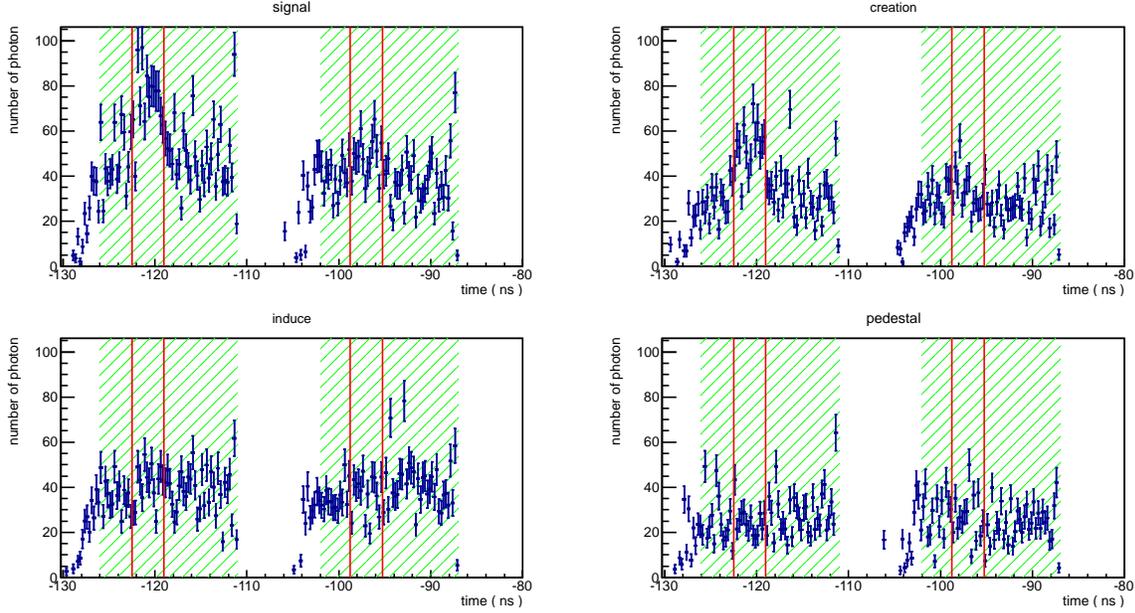}
\caption{
Charge sums within time windows of peak-like structures
in units of single-photon equivalent charge
that correspond to the numbers of photon-like signals,
as a function of the arrival time $t_0$ for the four trigger patterns
: signal, pedestal, green residual and red residual
within the time domains $T\{1\}$ and $T\{2\}$ with a 3.5~ns interval
subtended by the red vertical lines, respectively.
The green shaded bands indicate the online analysis windows of
15~ns time duration.
}
\label{Fig6}
\ecent
\end{figure}

Figure \ref{Fig6} shows 
charge sums within time windows of peak-like structures,
equivalently the numbers of photon-like signals,
as a function of the observed arrival time $t_0$ for the four trigger patterns.
The number of photons is counted in units of single-photon equivalent charge
within the time domains $T\{1\}$ and $T\{2\}$, respectively.
Table \ref{Tab1} summarizes the number photons in units of
single-photon equivalent charge in the four trigger patterns
in the time domains $T\{1\}$ and $T\{2\}$, respectively,
with the number of analyzed events in trigger patterns $i=S,C,I,P$,
respectively.

\begin{table}[]
\begin{tabular}{ c|c|c|c } \hline
Trigger $i$   & $n_{i1}$ & $n_{i2}$ & $W_i$           \\ \hline\hline
$S$           & 1036          & 675          & $5.97317383\times10^8$ \\ \hline
$P$           & 391           & 422          & $6.65181553\times10^8$ \\ \hline
$C$           & 822           & 513          & $6.65200665\times10^8$ \\ \hline
$I$           & 574           & 510          & $5.97300398\times10^8$ \\ \hline
\end{tabular} 
\caption{Observed raw numbers of photon-like signals $n_{ij}$ in two time domains
specified with the polarization states $j=1,2$ for each trigger 
pattern $i=$ $S$(signal), $P$(pedestal dominated by thermal noise), 
$C$(green laser), and $I$(red laser) where $W_i$ indicates the number 
of analyzed events in trigger pattern $i$.
}
\label{Tab1}
\end{table} 

\section{Results}
We take two steps to discuss the existence of signal photons
from the four-wave mixing process.
First, we investigate whether the acceptance-uncorrected
$N_{S1}$ and $N_{S2}$ 
indicate deviations from zero beyond the inclusive
errors in individual polarization paths.
We then set upper limits on the coupling-mass relation, 
if there is no statistically significant number of four-wave mixing signals.
Otherwise, we discuss the polarization dependence of the
observed finite numbers of four-wave mixing signals.

The acceptance-uncorrected numbers of photon-like signals $N_{S1}$ and $N_{S2}$ 
with signal triggers(S) after subtraction between four patterns of triggers
were obtained based on the relation in Eq.(\ref{eq_signal}) as follows:
\beqa\label{eq_raw}
N_{S1} = 75 \pm 51(stat.) \pm 96(syst. I) \pm 7(syst. II), \\ \nnb
N_{S2} = 83 \pm 44(stat.) \pm 96(syst. I) \pm 7(syst. II).
\eeqa
The statistical errors were calculated by taking propagation of statistical
errors associated with the subtraction using the photon numbers in 
Tab.\ref{Tab1} into account.
The systematic errors $I$ associated with the subtraction
between four trigger patterns were evaluated by focusing on the behavior 
of the numbers of photon-like signals in the time windows outside
$T\{1\}$ or $T\{2\}$ domains, 
denoting by $\overline{T\{1\}} \& \overline{T\{2\}}$,
where only unpolarized background photon-like signals are expected
\footnote{The residual of pulsed creation laser beam is only found either in
$T\{1\}$ or $T\{2\}$ domains, 
even though the polarization states can be distinguished by the two optical
fiber paths. 
On the other hand, the residual of the CW inducing laser beam can be found in
any time domains, however, the polarization states cannot be distinguished 
by the two optical fiber paths eventually, 
because photons even with one polarization state can 
produce signals in the two time domains equally as long as these
incident timings are randomly distributed.
Therefore, if $\overline{T\{1\}} \& \overline{T\{2\}}$ domain 
is chosen, no polarization dependence of the number of residual
photon-like signals is expected after all, as long as
the subtraction between four trigger patterns is ideally performed.
Needless to say, thermal noises should not have the polarization dependence.
}, hence,
the difference between $N_{S\overline{1}}$ and $N_{S\overline{2}}$ 
should not appear and the deviation from zero gives the systematic uncertainty
on the subtraction method.
We randomly combined 14 independent time bins of 0.25ns
interval among the $\overline{T\{1\}} \& \overline{T\{2\}}$ domain
and count the number of photon-like signals 
within the same time window as that of $T\{1\},T\{2\}$ domains, {\it i.e.},
3.5ns = 0.25ns $\times$ 14. 
The root-mean square of such counts eventually gives the systematic uncertainty 
on the baseline counts in $T\{1\}$ and $T\{2\}$ domains
associated with the subtraction method.
The systematic errors $II$ originates from the ambiguity on the offline cut
to define a peak to count the number of photon-like signals. 
These are estimated by changing the offline cut value
from 0.5 to 2.0 photon equivalent charge.
The numbers in Eq.(\ref{eq_raw}) indicate there is no significant four-wave mixing
signal within one standard deviation in the both polarization
states $\{1\}$ and $\{2\}$
\footnote{
The expected acceptance-uncorrected numbers of photons with \{2\}-state from 
atomic four-wave mixing processes are evaluated 
as $\sim 10^{-22}$ and $\sim 10^{-9}$ photons 
from the residual gas and optical elements for the same shot statistics
as the present search,
respectively, based on the supplementary measurements with higher intensity 
laser fields where four-wave mixing yields were evaluated 
as a function of residual gas pressures.
We estimated these numbers of photons by scaling the results in the
supplementary measurement down to the laser peak power of the two-color laser
fields with the same optical elements and focusing geometry 
as the present search by taking differences of detector acceptances and 
shot statistics between the present search and the supplementary measurements 
into account.
}.

In order to evaluate the acceptance-corrected photon numbers 
${\mathcal N}_{S1}$ and ${\mathcal N}_{S2}$,
we need to correct the bias due to the difference of detection
efficiencies in the two optical fibers in Fig.\ref{Fig3} 
for selecting $\{1\}$ and $\{2\}$ states, respectively. 
We thus parameterize the overall efficiencies, $\epsilon_{1,2}$ 
for $\{1\}$ and $\{2\}$ states, respectively, as
\beq\label{eq_efficiency}
\epsilon_1 \equiv \epsilon_{opt1}\epsilon_D, \qquad
\epsilon_2 \equiv \epsilon_{opt2}\epsilon_D
\eeq
where $\epsilon_{opt1,2}$ express 
optical collection efficiencies by the combination of the polarization 
beam splitter and optical fibers equipped with two lenses, and $\epsilon_D$
is the pure detection efficiency of the photomultiplier not including
the optical paths (see the enlarged SPD part in Fig.\ref{Fig3}). 
What we directly measure experimentally is a branching ratio between 
two paths containing exactly the same optical components and the detector
as those used for the pilot measurement,
which corresponds to the ratio of these two efficiencies
\beq\label{eq_branch}
B \equiv \frac{\epsilon_1}{\epsilon_2}.
\eeq
This quantity is determined by taking the ratio between the numbers of photons
in $T\{1\}$ and $T\{2\}$ at $\Theta=90$~deg. and $\Theta=0$~deg.,
respectively, in Fig.\ref{Fig3pol}.
With $B$ the acceptance-uncorrected photon numbers are expressed as
\beq\label{eq_M1M2}
{\mathcal N}_{S1} = \frac{N_{S1}}{\epsilon_1} = \frac{N_{S1}}{B\epsilon_2} 
= \frac{N_{S1}}{B\epsilon_{opt2}\epsilon_D}, \qquad
{\mathcal N}_{S2} = \frac{N_{S2}}{\epsilon_2} = \frac{N_{S2}}{\epsilon_{opt2}\epsilon_D},
\eeq
where the second of Eq.(\ref{eq_efficiency}) is substituted,
and $\epsilon_D$ is known by the other measurement in advance.

As the second step, 
we evaluate the upper limit on the coupling-mass relation, where 
we regard the acceptance-uncorrected 
uncertainty $\delta N_{S2}$ in the polarization state $\{2\}$ 
as the one standard deviation $\sigma$
in the following Gaussian type of distribution with the mean value 
$\mu = N_{S2}$ 
\beq
1-\alpha = \frac{1}{\sqrt{2\pi}\sigma}\int^{\mu+\delta}_{\mu-\delta}
e^{-(x-\mu)^2/(2\sigma^2)} dx = \mbox{erf}\left(\frac{\delta}{\sqrt 2 \sigma}\right),
\eeq
where the estimator $x$ corresponds to $N_{S2}$ in our case, and
the confidence level is given by $1-\alpha$\cite{PDGstatistics}.
As the upper limit estimate, we apply $2 \alpha = 0.05$ 
with $\delta = 2.24 \sigma$ to give a confidence level of $95\%$
in this analysis.

We then require that the expectation value of $N_{S2}$ 
coincides with $2.24\sigma$ to obtain the upper limit at 
the 95\% confidence level.
In order to relate the upper limit on $N_{S2}$ with the upper limit
on the coupling-mass relation directly, we further need to deduce ${\mathcal N}_{S2}$ 
corresponding to the unbiased number of signal photons immediately 
after the focal point via the second of Eq.(\ref{eq_M1M2})
\beq
{\mathcal N}_{S2} = \frac{N_{S2}}{\epsilon_{opt2} \epsilon_D} = 
\frac{2.24 \delta N_{S2}}{\epsilon_{opt2} \epsilon_D}.
\eeq

\begin{figure}
\bcent
\includegraphics[width=17.0cm]{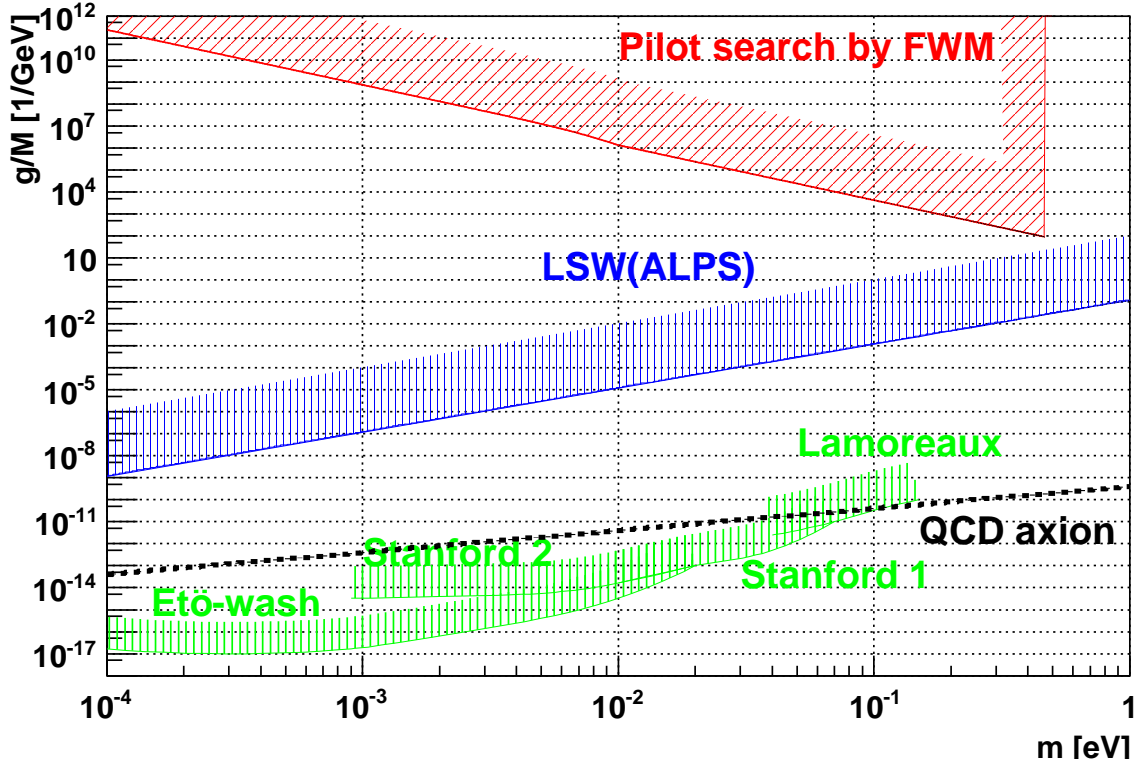}
\caption{
Upper limits on the coupling $g/M$ - mass $m$ relation for the scalar field exchange.
The excluded region by this pilot search based on four-wave mixing (FWM) in QPS 
is indicated by the slantly shaded area.
The vertically shaded areas show the excluded regions by the other scalar field searches:
the ALPS experiment~\cite{ALPS}( the sine function part of the sensitivity curve
is simplified by unity for the drawing purpose) which is
one of the "Light Shinig through a Wall" (LSW) experiments, searching for
non-Newtonian forces based on the torsion balance techniques
(Et\"{o}-wash\cite{Eto-wash}, Stanford1\cite{Stanford1}, Stanford2\cite{Stanford2})
and the Casimir force measurement (Lamoreaux\cite{Lamoreaux}).
As a reference for the future pseudoscalar searches, 
we also put the expected mass-coupling relation based on 
the QCD axion scenario for $E/N =0$~\cite{QCDaxions} (KSVZ model~\cite{KSVZ})
indicated by the inclining dotted line.
}
\label{Fig7}
\ecent
\end{figure}

Figure \ref{Fig7} shows the upper limit on the coupling-mass relation
at the 95\% confidence level which is calculated by requiring 
\beq\label{eq_WY}
W_s {\mathcal Y} = {\mathcal N}_{S2}
\eeq
in Eq.(\ref{eq24})
with the experimental parameters summarized in Tab.\ref{Tab1} and \ref{Tab2}.
The excluded domain by this pilot search based on four-wave mixing (FWM) 
in QPS is indicated by the slantly shaded area.
As references, we put existing upper limits by the other types of scalar field 
searches by vertically shaded areas: 
the ALPS experiment~\cite{ALPS}(the sine function part of the sensitivity curve
is simplified by unity for the drawing purpose) which is
one of the "Light Shinig through a Wall" (LSW) experiments, searching for
non-Newtonian forces based on the torsion balance techniques
(Et\"{o}-wash\cite{Eto-wash}, Stanford1\cite{Stanford1}, Stanford2\cite{Stanford2})
and the Casimir force measurement (Lamoreaux\cite{Lamoreaux}).
The domains below the vertically shaded areas are all excluded.
We note that if we require the proper
polarization combinations between initial and final states for
pseudoscalar fields~\cite{DEptep}, we are able to test the QCD axion models
in the near future. As a reference, we show the expected mass-coupling relation
based on the QCD axion scenario for $E/N =0$~\cite{QCDaxions}
(KSVZ model~\cite{KSVZ}) which is indicated by the inclining dotted line.

\section{Conclusion}
We demonstrated the pilot search for scalar fields in the mass region
$m \le 0.46$~eV via four-wave mixing of two-color laser fields. 
There is no significant four-wave mixing signals.
We obtained the upper limit on the coupling-mass relation 
at a confidence level of 95\%,
which approximately follows $m^{-5/2}$ scaling with the highest sensitivity
$g/M=92.6$~GeV${}^{-1}$ at $m = 0.46$~eV. 
The concept of the experimental setup as well as the analysis procedure
is extendable to higher intensity laser systems in the near future.

\section*{Appendix A: Re-evaluation of the signal yield }
Given the formulae in Refs.\cite{DEptp, DEapb, DEptep} and the
correction suggested by \cite{Correction}, 
we re-evaluate the signal yield with a more
direct formulation than that based on the concept of {\it cross section}.
The {\it cross section} is naturally applied to a beam flux normal to a target
or the extension to head-on particle colliders where all beams are on the same axis.
In QPS, however, this view point is not necessarily convenient
due to the wide distribution of tilted incident fluxes. In such a case,
we can adopt the more convenient formulation~\cite{BJ},
which is useful, for instance, to evaluate the number of interactions 
in plasma where the concept of {\it a beam flux to a target} is no longer clear.
With the notations in Fig.\ref{Fig2} and
the Lorentz invariant phase space factor $dL_{ips}$
\beq\label{eqLIPS}
dL_{ips} = (2\pi)^4 \delta(p_3+p_4-p_1-p_2)
\frac{d^3p_3}{2\omega_3(2\pi)^3}\frac{d^3p_4}{2\omega_4(2\pi)^3},
\eeq
the signal yield can be formulated as~\cite{BJ}
\beqa\label{eq_Y}
{\mathcal Y} &=& \int dt d\vec{r} \rho_1(\vec{r},t) \rho_2(\vec{r},t)
\int d\vartheta \frac{1}{2\omega 2\omega} \rho(\vartheta) |{\mathcal M(\vartheta)}|^2 dL_{ips}
\nnb\\
&\equiv& {\mathcal D}\left[s/L^3\right] \overline{\Sigma}\left[L^3/s\right],
\eeqa
where 
${\mathcal M(\vartheta)}$ is the Lorentz invariant transition amplitude 
as a function of incident angle $\vartheta$ with the normalized
statistical weight $\rho(\vartheta)$ due to the uncertainty of the
incident angle, ${\mathcal D}$ is time-integrated density and 
$\Sigma$ is the interaction volume per unit time. 
We note the dimensions explicitly with time $s$ and length $L$ in [ ]
of Eq.(\ref{eq_Y}). 
On the other hand, it may be possible to factorize 
the yield based on the concept of
{\it time-integrated luminosity} $\times$ {\it cross section}
as follows
\beqa\label{eqYsigma}
{\mathcal Y} &=& 
\int d\vartheta \rho(\vartheta)
\left( \int dt d\vec{r} \rho_1(\vec{r},t) \rho_2(\vec{r},t) K(\vartheta) \right)
\left(
\frac{1}{K(\vartheta) 2\omega 2\omega} 
|{\mathcal M(\vartheta)}|^2 dL_{ips}
\right)
\nnb\\
&\equiv& 
\int d\vartheta \rho(\vartheta) {\mathcal L(\vartheta)}\left[s/(L^2 \cdot s)\right]
\sigma(\vartheta)\left[L^2\right],
\eeqa
where $K$ corresponds to the relative velocity of incoming particle beams
for a given $\vartheta$
based on the M$\o$ller's Lorentz invariant factor~\cite{Moeller}. 
The relative velocity $K$ is defined as~\cite{K-factor}
\beq\label{eqK}
K(\vartheta) \equiv
\sqrt{(\vec{v_1}-\vec{v_2})^2 -
\frac{(\vec{v_1} \times \vec{v_2})^2}{c^2}}
= 2c\sin^2\vartheta,
\eeq
with the notation in Fig.\ref{Fig2}.
If $\rho(\vartheta)$ is a mere number, which is the normal case 
for collisions with two independent beams with a fixed relative 
velocity, this factorization is robust. In the case of the photon QPS,
however, the concept of relative velocity between a pair of incident photons
could be ambiguous due to spreads of single photon wavefunctions 
near the waist\footnote{Our attitude might be strongly biased by the Copenhagen
interpretation. If pre-existing momenta in advance of
measurement could be realized in nature, 
it might allow us to explicitly define relative velocity.
The approach expressed in Eq.(\ref{eq_Y}), however,
can circumvent this vague point originating from
the different interpretations on {\it realism} in quantum mechanics.},
while the squared scattering amplitude itself has to be averaged
over a possible range on $\vartheta$ due to that unavoidable uncertainty.
In order to recourse to the concept of a cross section, however,
we need to choose a proper relative velocity value to convert a
squared scattering amplitude into a cross section.
We then face difficulty in uniquely determining a $K$-factor 
with respect to the averaged squared scattering amplitude over 
a range of $\vartheta$ in QPS. 
The formulation in Eq.(\ref{eq_Y}) without recourse to the concept
of a cross section, on the other hand, ambiguity originating from 
choices of a proper $K$-factor is all avoidable.
Therefore, we re-formulate the signal yield based on Eq.(\ref{eq_Y}) 
as follows.

We first express the squared scattering amplitude
for the case when a low-mass field is exchanged
in the s-channel via a resonance state with the symbol to describe
polarization combinations of initial and final states $S$.
\beq\label{eq_M2}
|{\mathcal M}_S|^2 \approx  (4\pi)^2 \frac{a^2}{\chi^2+a^2},
\eeq
where
$\chi =\omega^2 -\omega_r^2$
with the resonance condition $m = 2\omega_r\sin\vartheta_r$
for a given mass $m$ and $a$ is expressed as
\beq\label{eq_a}
a = \frac{\omega^2_r}{8\pi}\left(\frac{g m}{M}\right)^2
=\frac{m \Gamma}{2\sin^2\vartheta_r}
\eeq
with the resonance decay rate of the low-mass field
\footnote{
We note that if $M \sim M_{Planck} \sim 10^{18}$ GeV and 
$g \sim \alpha_{qed} \sim 10^{-2}$,
the decay rate for $m \sim 10^{-9}$~GeV becomes $\sim 10^{-41}$ Hz, 
hence, the lifetime is quite long.
Therefore, we need to stimulate the decay at the same time as the production
by supplying an inducing laser field in advance of the interaction,
that is, we treat the interaction as an instantaneous scattering process 
where the production and the decay cannot be distinguished in spacetime.
}
\beq
\Gamma=(16\pi)^{-1} \left( g M^{-1}\right)^2 m^3.
\label{mxelm_4a}
\eeq

The resonance condition is satisfied when the center-of-mass system (CMS)
energy between incident two photons $E_{CMS}=2\omega\sin\vartheta$
coincides with the given mass $m$.
At a focused geometry of an incident laser beam, however,
$E_{CMS}$ cannot be uniquely specified due to the momentum uncertainty
of incident waves. Although the incident laser energy has the intrinsic
uncertainty, the momentum uncertainty
or the angular uncertainty between a pair of incident photons dominates
that of the incident energy.
Therefore, we consider the case where only angles of incidence
$\vartheta$ between randomly chosen pairs of photons
are uncertain within $0<\vartheta\le\Delta\vartheta$
for a given focusing parameter by fixing the incident energy. The treatment
for the intrinsic energy uncertainty is explained in Appendix B later.
We fix the laser energy $\omega$ at the optical wavelength
\beqa\label{exeq_13}
\omega_{opt}^2 = \frac{m^2}{4\vartheta^2_r} \sim 1\mbox{eV}^2,
\eeqa
while the resonance condition depends on the incident angle uncertainty.
This gives the expression for $\chi$ as a function of $\vartheta$
\beqa\label{exeq_14}
\chi(\vartheta) = w^2_{opt} - w^2_r(\vartheta)
= \frac{m^2}{4\vartheta^2_r} - \frac{m^2}{4\vartheta^2}
= \left(1-(\vartheta_r/\vartheta)^2\right) \omega_{opt}^2,
\eeqa
where 
\beq\label{eq_dvartheta}
d\vartheta = \frac{\vartheta_r}{2\omega^2_{opt}} (1-\frac{\chi}{\omega^2_{opt}})^{-3/2} d\chi.
\eeq
We thus introduce the averaging process for
the squared amplitude $\overline{|{\mathcal M}_S|^2}$
over the possible uncertainty on incident angles
\beqa\label{eq_M2average}
\overline{|{\mathcal M}_S|^2} = \int^{\pi/2}_{0} \rho(\vartheta)
|{\mathcal M}_S(\vartheta)|^2 d\vartheta
\eeqa
where
${\mathcal M}_S$ specified with a set of physical parameters $m$ and $gM^{-1}$
is expressed as a function of $\vartheta$, and
$\rho(\vartheta)$ is the probability distribution function
as a function of the uncertainty on $\vartheta$ within an incident pulse.

We review the expression for the electric field of the Gaussian laser
propagating along the $z$-direction
in spatial coordinates $(x,y,z)$~\cite{Yariv} as follows:
\beqa\label{eq_Gauss}
\vec{E}(x,y,z) = \vec{E_0}
\qquad \qquad \qquad \qquad \qquad \qquad \qquad \qquad \qquad \nnb\\
\frac{w_0}{w(z)}\exp
\left\{
-i[kz-H(z)] - r^2 \left( \frac{1}{{w(z)}^2}+\frac{ik}{2R(z)} \right)
\right\},
\eeqa
%
where $E_0$ electric field amplitude, 
$k=2\pi/\lambda$, $r=\sqrt{x^2+y^2}$, $w_0$ is the minimum waist,
which cannot be smaller than $\lambda$ due to the diffraction limit, and
other definitions are as follows:
%
\beqa\label{eq_wz}
{w(z)}^2 = {w_0}^2
\left(
1+\frac{z^2}{{z_R}^2}
\right),
\eeqa
\beqa\label{eq_Rz}
R = z
\left(
1+\frac{{z_R}^2}{z^2}
\right),
\eeqa
\beqa\label{eq_etaz}
H(z) = \tan^{-1}
\left(
\frac{z}{z_R}
\right),
\eeqa
\beqa\label{eq_zr}
z_R \equiv \frac{\pi{w_0}^2}{\lambda}.
\eeqa

With $\theta$ being an incident angle of a single photon in the Gaussian beam,
the angular distribution $g(\theta)$ can be approximated as
\beq\label{eqGaussAngular}
g(\theta) \sim \frac{1}{\sqrt{2\pi}\Delta\theta} 
\exp\left\{-\frac{\theta^2}{2\Delta\theta^2}\right\},
\eeq
where the incident angle uncertainty in the Gaussian beam
$\Delta\theta$ is introduced within the physical range $|\theta|<\pi/2$ as
\beqa\label{eq15}
\Delta\theta 
\sim 
\frac{\lambda_c}{\pi w_0} 
= \frac{d}{2f},
\eeqa
with the wavelength of the creation laser $\lambda_c$, the beam diameter $d$, 
the focal length $f$, and the beam waist $w_0 = \frac{f\lambda_c}{\pi d/2}$
as illustrated in Fig.\ref{Fig1}.
For a pair of photons 1, 2 each of which follows $g(\theta)$, the incident
angle between them is defined as
\beq\label{eqVartheta}
\vartheta = \frac{1}{2}|{\theta_1 - \theta_2}|.
\eeq
With the variance $\Delta\vartheta^2 = 2(\frac{1}{4}\Delta\theta^2)$,
the pair angular distribution $\rho(\vartheta)$ is then approximated as
\beq\label{eq_rho}
\rho(\vartheta) \sim \frac{2}{\sqrt{\pi}\Delta\theta} 
\exp\left\{
-\left(\frac{\vartheta}{\Delta\theta}\right)^2
\right\}
\sim \frac{2}{\sqrt{\pi}\Delta\theta}
\quad \mbox{for} \quad 0<\vartheta<\pi/2
\eeq
where 
the coefficient 2 of the amplitude is caused by limiting $\vartheta$ 
to the range $0 < \vartheta < \pi/2$,
and $\left(\frac{\vartheta}{\Delta\theta}\right)^2 \ll 1$
is taken into account because $\Delta\theta$ in Eq.(\ref{eq15})
also corresponds to the upper limit by
the focusing lens based on geometric optics. 
This distribution is consistent with the flat top distribution
applied to Ref.\cite{DEapb, DEptep} except the coefficient. 

We now re-express the average of the squared scattering amplitude as
a function of $\chi \equiv a\xi$
in units of the width of the Breit-Wigner(BW) distribution $a$
by substituting Eq.(\ref{eq_M2}) and (\ref{eq_rho}) 
into Eq.(\ref{eq_M2average}) with Eq.(\ref{eq_dvartheta})
\beqa\label{eq_M2a}
\overline{|{\mathcal M}_S|^2} 
=
\frac{(4\pi)^2}{\sqrt{\pi} \omega^2_{opt}}
\left(\frac{\vartheta_r}{\Delta\theta}\right) a
{\mathcal W},
\eeqa
where we introduce the following constant
\beq\label{eq_calW}
{\mathcal W} \equiv
\int^{\frac{\omega^2_{opt}}{a} \{1-(\vartheta_r/(\pi/2))^2 \}}_{-\infty}
W(\xi) \frac{1}{\xi^2+1} d\xi
\eeq
with 
\beq\label{eq_W}
W(\xi) \equiv (1-\frac{a\xi}{\omega^2_{opt}})^{-3/2}.
\eeq
In Eq.(\ref{eq_calW})
the weight function $W(\xi)$ is the positive and monotonic function within
the integral range and the second term is the Breit-Wigner(BW) function with 
the width of unity.
Note that $\overline{|{\mathcal M}_S|^2}$ is now explicitly proportional to $a$
but not $a^2$. This gives the enhancement factor $a$ compared to the case
$\overline{|{\mathcal M}_S|^2} \propto a^2$ where no resonance state is contained
in the integral range controlled by $\Delta\theta$ experimentally.
The integrated value of the pure BW function
from $\xi = -1$ to $\xi = +1$ gives $\pi/2$, while that
from $\xi = -\infty$ to $\xi = +\infty$ gives $\pi$.
The difference is only a factor of two.
The weight function $W(\xi)$ of the kernel is almost unity for small $a\xi$,
that is, when $a$ is small enough with a small mass and a weak coupling.
For instance, in the coupling-mass range covered by Fig.\ref{Fig7}, 
$g/M \sim 10^{11}$~GeV${}^{-1}$ and $m \sim 10^{-13}$~GeV gives $a \sim 10^{-5}$,
while $g/M \sim 10^2$~GeV${}^{-1}$ and $m \sim 10^{-10}$~GeV gives $a \sim 10^{-17}$.
In such cases the integrated value of Eq.(\ref{eq_calW}) is close to that of BW,
because the weight function is close to unity and also the upper limit of the
integral range in Eq.(\ref{eq_calW}) is large 
for $\vartheta_r/\Delta\vartheta < 1$ by the $a^{-1}$ dependence.
In Ref.\cite{DEptep}, we thus approximate ${\mathcal W}$ as $\pi/2$ for the
conservative estimate.

Let us remind of the partially integrated cross section over the solid angle
of the signal photon $\omega_3$
which corresponds to Eq.(11) in Ref.\cite{DEptep}. The expression
before taking the average over $\vartheta$, hence as a function of $\vartheta$, 
is as follows
\beq\label{eq_intgsigma}
\sigma(\vartheta) =
\frac{{\mathcal F_S} |{\mathcal M}_S(\vartheta)|^2}{(8\pi\omega)^2 \sin^4\vartheta}
\int^{\overline{\theta_3}}_{\underline{\theta_3}} 
\left(\frac{\omega_3}{2\omega}\right)^2 \sin\theta_3 d\theta_3.
\eeq
We then convert the interaction cross section $\sigma(\vartheta)$
into the interaction volume per unit time $\Sigma(\vartheta)$
by simply multiplying the relative velocity $K(\vartheta)=2c\sin^2\vartheta$ 
\beq\label{eq_intgvol}
\Sigma(\vartheta) =
\frac{{2c \mathcal F_S} |{\mathcal M}_S(\vartheta)|^2}{(8\pi\omega)^2 \sin^2\vartheta}
\int^{\overline{\theta_3}}_{\underline{\theta_3}} 
\left(\frac{\omega_3}{2\omega}\right)^2 \sin\theta_3 d\theta_3
\sim 
\frac{{2c^3 \mathcal F_S} |{\mathcal M}_S(\vartheta)|^2}{(8\pi)^2} 
\left(\frac{\lambda_c}{2\pi c}\right)^2 \frac{\delta u}{4}
\quad \left[L^3/s\right]
\eeq
where the creation laser wavelength $\lambda_c$,
$\delta u \equiv \overline{u} - \underline{u}$ with approximations $u^2 \ll 1$
and $\vartheta \ll 1$,
and $c$ and $\hbar$ are restored to confirm the dimension explicitly.
The averaged value over $\vartheta$ is then expressed as
\beqa\label{eq_Sigma}
\overline{\Sigma} &=& \int_0^{\pi/2} d\vartheta \rho(\vartheta) \Sigma(\vartheta)
= \frac{{c \mathcal F_S} \lambda^2_c \delta u }{2(16\pi^2)^2} 
\int_0^{\pi/2} d\vartheta \rho(\vartheta) |{\mathcal M}_S(\vartheta)|^2
\nnb\\
&\sim&
\frac{{c \mathcal F_S} \lambda^2_c \delta u }{2(16\pi^2)^2} 
\frac{(4\pi)^2}{\sqrt{\pi} \omega^2_{opt}}
\left(\frac{\vartheta_r}{\Delta\theta}\right) a {\mathcal W}
=
\frac{{c {\mathcal W}\mathcal F_S} \lambda^2_c \delta u }{2\sqrt{\pi}(4\pi)^2\omega^2_{opt}}
\left(\frac{\vartheta_r}{\Delta\theta}\right) a 
\nnb\\
&=&
\frac{{c {\mathcal W}\mathcal F_S} \lambda^2_c \delta u }{2\sqrt{\pi}(4\pi)^2\omega^2_{opt}}
\left(\frac{\vartheta_r}{\Delta\theta}\right)
\frac{\omega^2_r}{8\pi}\left(\frac{g m}{M}\right)^2
\nnb\\
&=&
\frac{{c {\mathcal W}\mathcal F_S} \lambda^2_c \delta u }{4\sqrt{\pi}(4\pi)^3}
\left(\frac{\vartheta_r}{\Delta\theta}\right) 
\left(\frac{g m}{M}\right)^2,
\eeqa
where Eq.(\ref{eq_M2a}) and $a$ in Eq.(\ref{eq_a}) are substituted 
in the second and third lines, respectively, and
$\omega_r \equiv \omega_{opt}$ via Eq.(\ref{exeq_13})
is identified for the last line.

We now consider the time-integrated density factor ${\mathcal D}$
in Eq.(\ref{eq_Y}) applicable to free propagation 
of the Gaussian laser beam for creation as illustrated in Fig.\ref{Fig1}. 
We first parameterize the density profile of an incident Gaussian laser beam 
being the focal point at $z=0$ with the pulse duration time
$\tau_c$ propagating over the focal length $f$ along $z$-axis as~\cite{Chi3Gas}
\beq\label{eq_GaussIntensity}
\rho_c(x,y,z,t) = \frac{2N_c}{\pi w^2(z)} \exp\left\{-2\frac{x^2+y^2}{w^2(z)}\right\}
\frac{1}{\sqrt{\pi}c\tau_c} \exp\left\{- \left( \frac{z^{'}}{c\tau_c} \right)^2\right\},
\eeq
where the central position of the creation pulse in $z$ coordinate is traced 
by the relation $z=ct$ as a function of
time $t$ while the pulse duration along $z$-direction 
is expressed with the local coordinate $z^{'}$, hence, $z^{'} = z-ct$,
and $N_c$ is the average number of creation photons per pulse.
Using the squared expression
\beq\label{eq_GaussIntensity2}
\rho^2_c(x,y,z,t) = \frac{4{N_c}^2}{\pi^3 (c\tau_c)^2} 
\frac{1}{w^4(z)}
\exp\left\{-4\frac{x^2+y^2}{w^2(z)}\right\}
\exp\left\{-2 \left( \frac{z^{'}}{c\tau_c} \right)^2\right\}
\eeq
based on Eq.(\ref{eq_Y}), ${\mathcal D}_c$ is then expressed as
\beqa\label{eq_Dc}
{\mathcal D}_c &=& \int^0_{-f/c} dt 
\int^{\infty}_{-\infty} dx
\int^{\infty}_{-\infty} dy
\int^{\infty}_{-\infty} dz
{\rho_c}^2(x,y,z,t)
\nnb\\
&=&
\frac{4N^2_c}{\pi^3(c\tau_c)^2} 
\int^0_{-f/c} dt \frac{1}{w^4(ct)} 
\sqrt{\frac{\pi w^2(ct)}{4}}
\sqrt{\frac{\pi w^2(ct)}{4}}
\int^{\infty}_{-\infty} dz^{'} \exp\left\{-2\left(\frac{z^{'}}{c\tau_c}\right)^2\right\}
\nnb\\
&=&
\frac{N^2_c}{\sqrt{2\pi}} \frac{1}{\pi{w_0}^2} \frac{1}{c\tau_c}
\int^0_{-f/c} dt \frac{1}{1+(ct/z_R)^2} 
\nnb\\
&=&
\frac{N^2_c}{\sqrt{2\pi}} \frac{1}{\pi{w_0}^2} \frac{1}{c\tau_c}
\frac{z_R}{c}\tan^{-1}\left(\frac{f}{z_R}\right) 
\quad \left[ s/L^3 \right],
\eeqa
where the time integration is performed during the pulse propagation
from $z=-f$ to $z=0$, because the photon-photon scattering never take place
when two photons in a pair are apart from each other at $z>0$.

We now evaluate the effect of the inducing laser beam.
The inducing effect is expected only when $p_4$ as a result of
scattering coincides with a photon momentum included
in the coherent state of the inducing beam.
In order to characterize $p_4$ from the interaction,
we first summarize the kinematic relations specified in Fig.\ref{Fig2}.
Although the CMS-energy varies depending on the incident angle $\vartheta$,
the interaction rate is dominated at the CMS-energy satisfying the resonance
condition $E_{CMS} = m = 2\omega\sin\vartheta_r$ 
via the Breit-Wigner weighting. Therefore, $p_4$ is essentially expressed
with the condition $\vartheta = \vartheta_r$ in Eq.(\ref{eq004}).
With $\vartheta_r \ll 1$, we obtain following relations
\beq\label{eq_theta4}
\theta_4 \sim {\mathcal R}^{-1/2}\vartheta_r \quad \mbox{ and } \quad
\theta_3 \sim {\mathcal R}^{1/2}\vartheta_r.
\eeq
These relations imply that the spontaneous interaction causes ring-like
patterns of emitted photons 
with cone angles $\theta_3$ and $\theta_4$ commonly constrained by
$\vartheta_r$, because the way to take a reaction plane determined 
by $p_{1,2,3,4}$ is symmetric around the $z$-axis.

We then evaluate how much fraction of photon momenta in
the inducing beam overlaps with the $p_4$, in other words, the
acceptance factor applied to the ideal Gaussian laser beam,
which was discussed within the plane wave approximation in \cite{DEptep}.
In the beam waist at $z=0$,
the angular spectrum representation gives the following electric field distribution
as a function of wave vector components $k_x$ and $k_y$ on the transverse plane $z=0$
based on Fourier transform of $\vec{E}(x,y,z=0)$ \cite{AngularRep}
\beq\label{eq_kT}
\vec{E}(k_x, k_y; z = 0 ) = 
\vec{E_0} \frac{w^2_0}{4\pi} e^{-\frac{w^2_0}{4}(k^2_x+k^2_y)} \equiv
\vec{E_0} \frac{w^2_0}{4\pi} e^{-\frac{w^2_0}{4}k^2_T},
\eeq
where the transverse wave vector component $k^2_T = k^2_x + k^2_y$
is introduced. With the paraxial approximation
$k_z \sim k - k^2_T/2k$, we can derive exactly the same expression 
for $E(x,y,z)$ in Eq.(\ref{eq_Gauss}) 
starting from this representation\cite{AngularRep}.
We apply Eq.(\ref{eq_kT}) to the inducing beam represented by the subscript 4
with $k_{T4} \equiv k_4\sin\theta_4 \sim k_4{\mathcal R}^{-1/2}\vartheta_r$
using the first in Eq.(\ref{eq_theta4})
and $w_{04} \equiv \frac{f\lambda_i}{\pi d/2}$.
For a range from $\underline{k_{T4}}$ to $\overline{k_{T4}}$
by denoting the underline and overline as the lower and upper values
on the corresponding variables, the acceptance factor ${\mathcal A}_i$
is estimated as
\beqa\label{eq_A4}
{\mathcal A}_i &\equiv&
\frac
{\int_{\underline{k_{T4}}}^{\overline{k_{T4}}} 2\pi k_{T4} E^2_4 dk_{T4}}
{\int_{0}^{\infty} 2\pi k_{T4} E^2_4 dk_{T4}}
= e^{-\frac{w^2_{04}}{2} \underline{k_{T4}}^2}
- e^{-\frac{w^2_{04}}{2} \overline{k_{T4}}^2}
\sim \frac{w^2_{04}}{2} (\overline{k_{T4}}^2-\underline{k_{T4}}^2) \nnb\\
&\sim& 
\frac{w^2_{04} k^2_4}{2}\vartheta^2_r
(\underline{\mathcal R}^{-1} - \overline{\mathcal R}^{-1}) \nnb\\
&\sim& 
2(\underline{\mathcal R}^{-1} - \overline{\mathcal R}^{-1})
\left( \frac{\vartheta_r}{\Delta\theta} \right)^2
= 
4\delta{\mathcal U}
\left( \frac{\vartheta_r}{\Delta\theta} \right)^2,
\eeqa
where the approximation in the first line is based on $w_{04} k_{T4} \ll 1$,
$k_{T4} \sim k_4{\mathcal R}^{-1/2}\vartheta_r$ is substituted in the second line,
and $w_{04}k_4 = \frac{2\pi w_{04}}{\lambda_i} \equiv 2/\Delta\theta_4$ with
$\Delta\theta_4 \sim \Delta\theta$ due to
the common focusing geometry $d$ and $f$ to those of the creation beam
is substituted in the third line by defining $\delta{\mathcal U} \equiv
\frac{\overline{u}-\underline{u}}{\overline{u}\underline{u}}$.
We note that Eq.(\ref{eq_A4}) now has the quadratic dependence on $\vartheta_r$
while the corresponding acceptance factor in \cite{DEptep} was proportional
to $\vartheta_r$ due to the plane wave approximation.

Because of the common optical element sharing the same optical axis, in advance,
the spatial overlap between creation and inducing lasers is satisfied in QPS.
In actual experiments, however, it is likely that the time durations
between the two laser pulses are prepared as $\tau_i \ge \tau_c$
for inducing $i$ and creation $c$ beams, respectively,
because shortness of $\tau_c$ is more important to enhances 
${\mathcal D}_c$ via the quadratic nature on the pulse energy.
Therefore, we further introduce a factor representing the spacetime 
overlap by assuming the pulse peaks in spacetime coincide with each other. 
Hence, the entire inducing effect is expressed as
\beq\label{eq_I}
{\mathcal I } = {\mathcal A_i} \left(\frac{\tau_c}{\tau_i}\right) N_i
= 
4\delta{\mathcal U}
\left( \frac{\vartheta_r}{\Delta\theta} \right)^2
\left(\frac{\tau_c}{\tau_i}\right) N_i.
\eeq

Therefore, the overall density factor 
${\mathcal D}_{c+i}$ including the inducing laser effect is expressed as  
\beqa\label{eq_Dciall}
{\mathcal D}_{c+i} &\sim& 
C_{mb} {\mathcal D}_c {\mathcal I} \nnb\\
&=&
C_{mb} \frac{N^2_c}{\sqrt{2\pi}} \frac{1}{\pi{w_0}^2} \frac{1}{c\tau_c}
\frac{z_R}{c}\tan^{-1}\left(\frac{f}{z_R}\right)
4\delta{\mathcal U} \left(\frac{\vartheta_r}{\Delta\theta}\right)^2
\left(\frac{\tau_c}{\tau_i}\right) N_i
\nnb\\
&=&
\frac{4}{\sqrt{2\pi}}
\frac{z_R}{\pi{w_0}^2 c^2 \tau_c}
\left(\frac{\tau_c}{\tau_i}\right)
\tan^{-1}\left(\frac{f}{z_R}\right)
\left(\frac{\vartheta_r}{\Delta\theta}\right)^2
\delta{\mathcal U} C_{mb} N^2_c N_i 
\nnb\\
&=&
\frac{4}{\sqrt{2\pi}}
\frac{1}{c^2 \lambda_c \tau_c}
\left(\frac{\tau_c}{\tau_i}\right)
\tan^{-1}\left(\frac{\pi d^2}{4\ f\lambda_c}\right)
\left(\frac{\vartheta_r}{\Delta\theta}\right)^2
\delta {\mathcal U} C_{mb} N^2_c N_i 
\eeqa
where 
$C_{mb}=1/2$ corresponds to the combinatorial factor 
originating from the choice of a frequency among the frequency multimode 
states in creation and inducing lasers as discussed in Ref.\cite{DEptep}.

Based on Eq.(\ref{eq_Y}) with Eqs.(\ref{eq_Sigma}) and (\ref{eq_Dciall}), 
the re-evaluated signal yield ${\mathcal Y}$ is finally expressed as
\beqa\label{eqY}
{\mathcal Y} &=& {\mathcal D}_{c+i} \overline{\Sigma}
\nnb\\
&=& 
\frac{4}{\sqrt{2\pi}}
\frac{1}{c^2 \lambda_c \tau_c}
\left(\frac{\tau_c}{\tau_i}\right)
\tan^{-1}\left(\frac{\pi d^2}{4\ f\lambda_c}\right)
\left(\frac{\vartheta_r}{\Delta\theta}\right)^2
\delta {\mathcal U} C_{mb} N^2_c N_i
\frac{{c {\mathcal W}\mathcal F_S} \lambda^2_c \delta u }{4\sqrt{\pi}(4\pi)^3}
\left(\frac{\vartheta_r}{\Delta\theta}\right) 
\left(\frac{g m}{M}\right)^2
\nnb\\
&=&
\frac{1}{\sqrt{2} \pi (4\pi)^3}
\left(\frac{\lambda_c}{c\tau_c}\right)
\left(\frac{\tau_c}{\tau_i}\right)
\tan^{-1}\left(\frac{\pi d^2}{4\ f\lambda_c}\right)
\left(\frac{\vartheta_r}{\Delta\theta}\right)^3
\delta u \delta {\mathcal U} 
\left(\frac{g m}{M}\right)^2
{\mathcal W}{\mathcal F_S} C_{mb} N^2_c N_i
\nnb\\
&=&
\frac{1}{\sqrt{2} \pi (4\pi)^3}
\left(\frac{\lambda_c}{c\tau_c}\right)
\left(\frac{\tau_c}{\tau_i}\right)
\Delta\theta^{-3}
\tan^{-1}\left(\frac{\pi d^2}{4\ f\lambda_c}\right)
\delta u \delta {\mathcal U} 
\left(\frac{gm}{M}\right)^2 \left(\frac{m}{2\omega{\mbox [eV]}}\right)^3
{\mathcal W}{\mathcal F_S} C_{mb} N^2_c N_i
\nnb\\
&\sim&
\frac{1}{64\sqrt{2}\pi^4}
\left(\frac{\lambda_c}{c\tau_c}\right)
\left(\frac{\tau_c}{\tau_i}\right)
\left(\frac{f}{d}\right)^3
\tan^{-1}\left(\frac{\pi d^2}{4\ f\lambda_c}\right)
\frac{(\overline{u}-\underline{u})^2}{\overline{u}\underline{u}}
\left(\frac{gm{\mbox[eV]}}{M{\mbox[eV]}}\right)^2 \left(\frac{m{\mbox[eV]}}{\omega{\mbox[eV]}}\right)^3
{\mathcal W}{\mathcal F_S} C_{mb} N^2_c N_i,
\nnb\\
\eeqa
where the parameters specified with [eV] apply natural units.
We note that the re-evaluated signal yield depends on
$m^5$ compared to the $m^2$ dependence in Ref.\cite{DEptep}. 
The additional $m^2$ dependence is caused by multiplying $K$-factor in
Eq.(\ref{eq_intgvol}) and the $m$ dependence is further added by Eq.(\ref{eq_A4}) due
to the ring-image acceptance for the inducing effect.
The sensitivity to lower mass domains thus diminishes.

\section*{Appendix B:
The effect of finite spectrum widths of
creation and inducing laser beams 
}
When creation and inducing lasers have finite spectrum widths, the effect
has impacts on the interaction rate of four-wave mixing as well as 
the spectrum width of the signal via energy-momentum conservation within 
the unavoidable uncertainty of photon momenta and energies in QPS. 
We have considered the effect of the finite line width of the inducing
laser field by assuming the line width of the creation
laser is negligibly small~\cite{DEptep}. In such a case, the energy uncertainties
in the process of four-wave mixing with the convention used in Fig.\ref{Fig2}
are described as
\beq\label{eq_widthptep}
\omega = \langle \omega \rangle,
\quad
\omega_4 = \langle \omega_4 \rangle \pm \delta\omega_4,
\quad \mbox{and} \quad
\omega_3 = \langle \omega_3 \rangle \pm \delta\omega_3
\eeq
for the creation laser, the inducing laser, and the signal, respectively, 
where $\langle \cdot\cdot \rangle$ denotes average of 
each spectrum, $\delta \omega_4$ expresses the intrinsic
line width caused by the energy uncertainty in the atomic
process to produce the inducing laser beam,
and $\delta\omega_3 = \delta\omega_4$ arises due to the energy conservation
\cite{DEptep}.

%
In this case $\overline{u}$ and $\underline{u}$ in Eq.(\ref{eqY}) via Eq.(\ref{eq000})
solely originates from the intrinsic line width of the inducing laser field.
However, this approximation is not valid for 
the case where the line width of the creation laser is equal to or wider 
than that of the inducing laser.
We now provide the prescription for the following general case:
\beq\label{eq_width}
\omega = \langle \omega \rangle + \delta\omega,
\quad
\omega_4 = \langle \omega_4 \rangle \pm \delta\omega_4,
\quad \mbox{and} \quad
\omega_3 = \langle \omega_3 \rangle \pm \delta\omega_3,
\eeq
where the intrinsic line width of the creation laser $\delta\omega$ is 
explicitly included.
In this case, at a glance,
Eq.(\ref{eq003}) must be modified so that a finite
net transverse momentum along the $x$-axis in Fig.\ref{Fig2} is 
introduced by the different incident photon energies
$\omega_1$ and $\omega_2$ with corresponding incident angles 
$\vartheta_1$ and $\vartheta_2$, respectively.
These notations are for a {\it nominal} laboratory frame defined 
by a pair of asymmetric incident photons, referred to as $L$-system. 
Here {\it nominal} implies 
that we cannot specify individual photon's incident wavelength 
as well as incident angle physically, that is,  
what we know {\it a priori} is only the ranges of uncertainties
on possible wavelengths and incident angles.
Exactly speaking,
all the calculations so far are based on the equal incident energy 
$\langle \omega \rangle$ with 
the equal incident angle $\vartheta$ in the averaged laboratory frame, 
referred to as
$\langle L \rangle$-system as illustrated in Fig.\ref{Fig2}, 
where the transverse momentum of the incident 
colliding system is zero on the average.
We regard the effect of these {\it nominally} possible $L$-systems 
as fluctuations around the averaged $\langle L \rangle$-system
We, therefore, attempt to transform $L$-systems into a 
$\langle L \rangle$-system so that $\delta\omega$ is effectively
invisible as follows
\beq\label{eq_widthAve}
\omega = \langle \omega \rangle,
\quad
\omega_4 = \langle \omega_4 \rangle \pm \Delta\omega_4,
\quad \mbox{and} \quad
\omega_3 = \langle \omega_3 \rangle \pm \Delta\omega_3,
\eeq
where $\delta\omega$ in Eq.(\ref{eq_width}) is absorbed 
into the effective line widths  of the final state photon energies in the
$\langle L \rangle$-system resulting in $\Delta\omega_4 > \delta\omega_4$
denoting $\Delta$ as the effective width in the $\langle L \rangle$-system
and accordingly $\Delta\omega_3 > \delta\omega_3$ arises
via energy-momentum conservation in that system.

As long as a resonance state with a finite mass is formed, a range
$0<\vartheta \le \Delta\vartheta$ in Fig.\ref{Fig1}
eventually contains the resonance angle
$\vartheta_r$ which satisfies the condition that the center-of-mass system
energy $E_{CMS}=2\langle\omega\rangle\sin\vartheta_r$ coincides with
the mass $m$ in the $\langle L \rangle$-system,
and any {\it nominal} $L$-systems which satisfy the resonance condition
can be generally formed by a transverse Lorentz boost of
the $\langle L \rangle$-system
with the relative velocity $\beta$ with respect to the velocity of light $c$.
Assigning the positive direction of $\beta$ to the
positive direction of $x$-axis in Fig.\ref{Fig2},
the energy and the transverse momentum relations between a $L$-system
and the $\langle L \rangle$-system are connected via~\cite{PDGkine}
\beqa\label{eq_boost}
\left( \begin{array}{cc} \omega_1\\ p_{1T}\\ \end{array} \right)
=
\gamma
\left(
\begin{array}{cc}
1 & -\beta \\
-\beta & 1\\
\end{array}
\right)
\left(
\begin{array}{cc}
\langle\omega\rangle \\
\langle\omega\rangle\sin\vartheta \\
\end{array}
\right),
\nnb\\
\left( \begin{array}{cc} \omega_2\\ p_{2T}\\ \end{array} \right)
=
\gamma
\left(
\begin{array}{cc}
1 & -\beta \\
-\beta & 1\\
\end{array}
\right)
\left(
\begin{array}{cc}
\langle\omega\rangle \\
-\langle\omega\rangle\sin\vartheta \\
\end{array}
\right)
\eeqa
with $\gamma \equiv (1-\beta^2)^{-1/2}$,
where the subscripts 1 and 2 correspond to the photon indices in Fig.\ref{Fig2},
respectively.
Since the relations of the energy components in Eq.(\ref{eq_boost}) indicate
that $p_1$ experiences frequency-down shift,
while $p_2$ does frequency-up shift,
we introduce another definitions of energy components of $p_1$ and $p_2$
after the transverse boost by
\beqa\label{eq_omega12}
\omega_1 = \gamma \langle\omega\rangle (1-\beta\sin\vartheta)
\equiv \langle\omega\rangle - \delta\omega
\equiv \langle\omega\rangle\left(1-\delta r\right), \nnb\\
\omega_2 = \gamma \langle\omega\rangle (1+\beta\sin\vartheta)
\equiv \langle\omega\rangle + \delta\omega
\equiv \langle\omega\rangle\left(1+\delta r\right)
\eeqa
where
a relative line width with respect to the mean energy of the creation laser
$\delta r \equiv \delta\omega / \langle\omega\rangle$
is implemented. We note the following physical and experimental conditions:
\beq\label{eq_condition}
0 < \beta < 1 \quad \mbox{and} \quad 0 < \delta r \ll 1,
\eeq
respectively. For convenience, we tentatively distinguish $\beta$
for $\omega_{1,2}$ by the subscripts in the following discussion.
 
As for $\omega_1$, from the first of Eq.(\ref{eq_omega12}) with $\delta r \ll 1$,
we obtain the following equation
\beqa\label{eq_betaT1quad}
(1+\sin^2\vartheta -2\delta r)\beta^2_{1} - 2\sin\vartheta\beta_{1} + 2\delta r = 0
\eeqa
with the solutions by assuming $\vartheta \ll 1$
\beqa\label{eq_betaT1}
\beta_{1\pm} \equiv \frac{\vartheta \pm \sqrt{D_1}}{1+\vartheta^2-2\delta r}
\sim (1-\vartheta^2+2\delta r)\left(\vartheta \pm \sqrt{D_1}\right),
\eeqa
where
\beq\label{eq_D1}
D_1 \sim \vartheta^2 - (1+\vartheta^2 - 2\delta r) 2\delta r
\sim \vartheta^2 - 2\delta r.
\eeq
Requiring $D_1 \ge 0$ gives a physical constraint
\beq\label{eq_condition}
2\delta r \le \vartheta^2.
\eeq
If this is satisfied,
$\sqrt{D_1} \sim (\vartheta^2 - 2\delta r)^{1/2}
\sim \vartheta - \delta r/\vartheta$ with $\delta r \ll 1$
gives
\beq\label{eq_deno1}
\vartheta\pm \sqrt{D_1} \sim \vartheta \pm
\left(\vartheta - \frac{\delta r}{\vartheta}\right).
\eeq
The double-sign symbol in Eq.(\ref{eq_betaT1}) gives following two solutions:
\beqa\label{eq_beta1pm}
\beta_{1+} &\sim& (1-\vartheta^2+2\delta r)
\left(2\vartheta-\frac{\delta r}{\vartheta}\right)
\sim 2\vartheta -\frac{\delta r}{\vartheta}, \nnb\\
\beta_{1-} &\sim& (1-\vartheta^2+2\delta r)
\left(\frac{\delta r}{\vartheta}\right)
\sim \frac{\delta r}{\vartheta} > 0,
\eeqa
where only $\beta_{1-}$ is physically acceptable,
while $\beta_{1+}$ is not, because the limit of
$\delta r \rightarrow 0$ must correspond to
$\beta = 0$ in our discussion.

Let us move on to discuss $\omega_2$.
From the second of Eq.(\ref{eq_omega12}) with $\delta r \ll 1$, we get
\beqa\label{eq_betaT2quad}
(1+\sin^2\vartheta +2\delta r)\beta^2_{2} + 2\sin\vartheta\beta_{2} - 2\delta r = 0
\eeqa
with the solutions by assuming $\vartheta \ll 1$
\beqa\label{eq_betaT2}
\beta_{2\pm} \equiv \frac{-\vartheta \pm \sqrt{D_2}}{1+\vartheta^2+2\delta r}
\sim (1-\vartheta^2-2\delta r)\left(-\vartheta \pm \sqrt{D_2}\right),
\eeqa
where
\beq\label{eq_D2}
D_2 \sim \vartheta^2 + (1+\vartheta^2 + 2\delta r) 2\delta r
\sim \vartheta^2 + 2\delta r>0.
\eeq
The approximation $\sqrt{D_2} \sim (\vartheta^2+2\delta r)^{1/2}
\sim \vartheta + \delta r/\vartheta$ with $\delta r \ll 1$
gives
\beq\label{eq_deno2}
-\vartheta\pm \sqrt{D_2} \sim -\vartheta \pm
\left(\vartheta + \frac{\delta r}{\vartheta}\right).
\eeq
The double-sign symbol in Eq.(\ref{eq_betaT2}) gives following two solutions:
\beqa\label{eq_beta2pm}
\beta_{2+} &\sim& (1+\vartheta^2+2\delta r)
\left(\frac{\delta r}{\vartheta}\right)
\sim \frac{\delta r}{\vartheta} > 0, \nnb\\
\beta_{2-} &\sim& (1+\vartheta^2+2\delta r)
\left(-2\vartheta-\frac{\delta r}{\vartheta}\right)
\sim -2\vartheta-\frac{\delta r}{\vartheta},
\eeqa
where only $\beta_{2+}$ is physically acceptable, while $\beta_{2-}$ is not
due to the same reason as $\omega_1$ in addition to the positivity condition.

A common $\beta$ is eventually determined as
\beq\label{eq_beta}
\beta \sim \beta_{1-} = \beta_{2+} = \frac{\delta r}{\vartheta}
\eeq
based on Eqs.(\ref{eq_beta1pm})
and (\ref{eq_beta2pm}).
By substituting $\delta r$ in Eq.(\ref{eq_beta}) into Eq.(\ref{eq_condition}),
the range of $\beta$ is expressed as
\beq\label{eq_betaTr}
\beta < \frac{\vartheta}{2}.
\eeq

Among $\vartheta$ within $\Delta \vartheta$,
$\vartheta_r$ is effectively enhanced based
on the Breit-Wigner distribution in the averaging process of the square
of the invariant scattering amplitude around
$E_{CMS}=2\sin\vartheta_r\langle\omega\rangle$ \cite{DEapb}.
Therefore, for a given mass
$m \sim 2\langle\omega\rangle\vartheta_r$ with $\vartheta_r \ll 1$,
the effective physical limit on $\beta$ is expressed as
\beq\label{eq_betaTr}
\beta_{r} \equiv \frac{\delta r}{\vartheta_r} < \frac{\vartheta_r}{2}.
\eeq
On the other hand, an instrumental full line width of a creation laser
$\Delta r \equiv \delta\omega_{full}/\langle\omega\rangle$
is given by the creation laser intrinsically. Therefore,
the range of $\beta$ can be maximally covered by
\beqa\label{eq_betaTc}
0 < \beta < \beta_{c} \equiv \frac{\Delta r}{\vartheta}
\eeqa
based on the relation in Eq.(\ref{eq_beta}).
If $\beta_c$ is smaller than $\beta_r$, however, the instrumental
condition limits $\beta$ to $\beta \le \beta_c$
before reaching the physical limit $\beta_r$ for a given mass parameter.
Therefore, we are required to choose smaller $\beta$,
either $\beta_{c}$ or $\beta_{r}$, depending on the relation
between an experimentally given line width of the
creation laser and a given mass parameter we search for.

Given a possible range of $\beta$ based on the relation between an intrinsic
line width of the creation laser and a given mass $m$,
we can construct a unique $\langle L \rangle$-system by inversely boosting
individual $L$-systems, where the intrinsic line width $\delta \omega_4$ is 
effectively broaden.
This is because a spectrum width is effectively embedded by the 
possible range of inverse boosts for a $\omega_4$ chosen among 
$\langle\omega_4\rangle - \delta\omega_4 \le \omega_4 \le
\langle\omega_4\rangle + \delta\omega_4$ in individual $L$-systems.
Therefore, combining the inverse-boost-originating spectrum width with the intrinsic 
line width of the inducing laser provides the effective inclusive range of $u$
defined in Eq.(\ref{eq000}) in the $\langle L \rangle$-system.

We first evaluate how much the common inverse boost originating
from the line width of the creation laser solely changes the range 
of $u$ from $\underline{u}_c$ to $\overline{u}_c$
in the $\langle L \rangle$-system for a $\omega_4$ within
the line width of the inducing laser field. By inversely applying 
the boost in Eq.(\ref{eq_boost}) to $p_4$ in Fig.\ref{Fig2},
we can express the upper and lower edges of the broadened energy range 
of $\omega_4$ with the approximation $\vartheta_r \ll 1$, respectively,
\footnote{
Note that the degree of freedom to flip $p_3$ and $p_4$ around the $z$-axis
by keeping the angular balance in Fig.1 results in
both frequency up and down shifts in $\omega_3$ and $\omega_4$.
}
\beqa\label{eq_uboost} 
\overline{u}_c\langle\omega\rangle
\sim l_i u\langle\omega\rangle
\gamma(1 + \beta\vartheta)
\equiv
l_i u\langle\omega\rangle
\min\{ \gamma_c(1 + \Delta r), \quad \gamma_r(1 + \frac{1}{2}\vartheta^2_r) \}
\nnb\\
\underline{u}_c\langle\omega\rangle
\sim l_i u\langle\omega\rangle
\gamma(1 - \beta\vartheta)
\equiv
l_i u\langle\omega\rangle
\min\{ \gamma_c(1 - \Delta r), \quad \gamma_r(1 - \frac{1}{2}\vartheta^2_r) \}
\eeqa
where 
$l_i$ represents the process to choose a $\omega_4$ 
within the relative line width $l_i \equiv 1 \pm \delta u_4/u$ 
with $\delta u_4 \equiv (\overline{u} - \underline{u})/2$ being 
the average value $\langle l_i\rangle=1$,
subscripts $c$ and $r$ in the right-hand side
denote the cases where $\beta_c$ and $\beta_r$ are used,
respectively, and min$\{A, B\}$ requires to choose smaller one 
between $A$ and $B$.

We then discuss the inclusive range of $u$ 
by combining the broadened $\omega_4$ in the $\langle L \rangle$-system 
with the intrinsic line width of the inducing laser.
We introduce a notation reflecting the combining process 
in the $\langle L \rangle$-system
\beq\label{eq_omega4L}
\omega_{<4>} \equiv l_c l_i u \langle \omega \rangle
\eeq
where 
$l_c \equiv 1 \pm \delta u_c/u$ 
with $\delta u_c \equiv (\overline{u_c} - \underline{u_c})/2$
being the average value $\langle l_c\rangle=1$,
is a coefficient describing the uncertainty by the inverse
transverse boost discussed above.
Since the intrinsic energy uncertainties of creation and inducing laser beams 
are independent,
the quadratic error propagation gives the following inclusive uncertainty on $\omega_4$
in the $\langle L \rangle$-system
\beq\label{eq_domega4L}
\delta^2 \omega_{<4>} = \left(\frac{\partial\omega_{<4>}}{\partial l_i}\right)^2\delta^2 l_i + 
\left(\frac{\partial\omega_{<4>}}{\partial l_c}\right)^2 \delta^2 l_c
= u^2 \langle \omega \rangle^2 \left[ 
\left(\frac{\delta u_4}{u}\right)^2 + \left(\frac{\delta u_c}{u}\right)^2 \right].
\eeq
Finally this gives the inclusive uncertainty on $u$
\beq\label{eq_du}
\delta u_{inc} \equiv \frac{\delta \omega_{<4>}}{\langle \omega \rangle } = 
u \sqrt{ \left(\frac{\delta u_4}{u}\right)^2 + \left(\frac{\delta u_c}{u}\right)^2 },
\eeq
and the upper and lower limits on $u$, which replace the limits in Eq.(\ref{eq24}) with
\beqa\label{eq_convou}
\overline{u} \equiv  u + \delta u_{inc} \mbox{ and}
\quad \underline{u} \equiv u - \delta u_{inc},
\eeqa
respectively, are obtained.

As a summary based on Eq.(\ref{eq_uboost}),
the effect of the line width of the creation laser
is less significant compared to that of the inducing laser
for a smaller mass range, while it has some impact
for a larger mass range, as long as the line width is comparable
to or larger than that of the inducing laser.

\section*{Acknowledgments}
K. Homma cordially thanks Y. Fujii for the long-term discussions
on the theoretical aspects relevant to the scalar field in the context of
cosmology. He has greatly benefited from valuable discussions
with S. Sakabe, M. Hashida and Y. Nakamiya.
He expresses his gratitude to T. Tajima and G. Mourou
for many aspects relevant to this subject.
This work was supported by the Grant-in-Aid for
Scientific Research no.24654069 and 25287060 from MEXT of Japan,
and also supported by MATSUO FOUNDATION.

%
%

%
%
\begin{table}[]
\begin{tabular}{ c|c } \hline
parameters   & values or comments           \\ \hline\hline
wavelength of creation laser $\lambda_c$    & 532 nm \\ \hline
relative line width of creation laser ($\delta\omega/\langle\omega\rangle$) & $4.6 \times10^{-6}$ \\ \hline
wavelength of inducing laser $\lambda_i$    & 633 nm \\ \hline
relative line width of inducing laser ($\delta\omega_4/\langle\omega_4\rangle$) & $3.2 \times 10^{-6}$ \\ \hline
$u = \omega_4/\omega$ based on Eqs.(\ref{eq000}) and (\ref{eq004})& 0.84 \\ \hline
focal length $f$                           & 200 mm \\ \hline
beam diameter of laser beams $d$           &  40 mm \\ \hline
upper mass range given by $\vartheta<\Delta\vartheta$ & 0.46 eV \\ \hline
azimuthal asymmetric factor ${\mathcal F}_S$ & ${\mathcal F}^{SC}_{1122}$ in Eq.(\ref{eq_Fs}) \\ \hline
duration time of creation laser pulse per injection $\tau_c$ & 0.75 ns \\ \hline 
duration time of inducing laser per injection $\tau_i$ (CW) & 1 s \\ \hline 
creation laser energy per $\tau_c$ & $(0.181 \pm 0.009)\mu$J \\ \hline
inducing laser energy per $\tau_i$ & $(1.87 \pm 0.19)$mJ \\ \hline
combinatorial factor in luminosity $C_{mb}$ & 1/2 \\ \hline
single photon detection efficiency $\epsilon_D$ & $(2.94 \pm 0.03)$\% \\ \hline
efficiency of the optical path 2 $\epsilon_{opt2}$ & $(4.3 \pm 0.1)$\% \\ \hline
branching ratio $B \equiv \epsilon_1/\epsilon_2$ & $0.824 \pm 0.002$ \\ \hline
$\delta N_{S2}$ in Eq.(\ref{eq_WY}) & 106 \\ \hline
\end{tabular} 
\caption{Experimental parameters used to obtain the upper limit on
the coupling-mass relation.}
\label{Tab2}
\end{table} 


\begin{references}
\bibitem{Axion}
R. D. Peccei and H. R. Quinn, Phys. Rev. Lett. 38, 1440 (1977);
S. Weinberg, Phys. Rev. Lett. 40, 223 (1978);
F. Wilczek, Phys. Rev. Lett. 40, 271 (1978).
\bibitem{STTL}
Y. Fujii and K. Maeda, {\it The Scalar-Tensor Theory of Gravitation}
Cambridge Univ. Press (2003);
Mark P. Hertzberg, Max Tegmark, and Frank Wilczek,
Phys. Rev. D {\bf 78}, 083507 (2008);
Olivier Wantz and E. P. S. Shellard, Phys. Rev. D {\bf 82}, 123508 (2010).
\bibitem{ISMD2011}
K. Homma, D. Habs, G. Mourou, H. Ruhl, and T. Tajima,
Prog. Theor. Phys. Suppl. No. 193, (2012).
\bibitem{TajimaHomma}
T. Tajima and K. Homma, Int. J. Mod. Phys. A vol. 27, No. 25, 1230027 (2012).
\bibitem{DEptp}
Y. Fujii and K. Homma,
Prog. Theor. Phys. 126: 531-553 (2011), arXiv:1006.1762 [gr-qc].
\bibitem{DEapb}
K. Homma, D. Habs, T. Tajima,
Appl. Phys. B 106:229-240 (2012),
(DOI: 10.1007/s00340-011-4567-3),arXiv:1103.1748 [hep-ph].
\bibitem{DEptep}
K. Homma, Prog. Theor. Exp. Phys. 04D004 (2012).
\bibitem{FWM}
S. A. J. Druet and J.-P. E. Taran, Prog. Quant. Electr. 7, 1 (1981).
\bibitem{Yariv}
Amnon Yariv, {\it Optical Electronics in Modern Communications}
Oxford University Press (1997).
\bibitem{FWMqed}
F. Moulin and D. Bernard, Opt. Commun. 164, 137 (1999);
E. Lundstr\"{o}m et al., Phys. Rev. Lett. 96, 083602 (2006);
J. Lundin et al., Phys. Rev. A 74, 043821 (2006);
D. Bernard et al., Eur. Phys. J. D 10, 141 (2000).
\bibitem{FifthForce}
Y. Fujii, Nature Phys. Sci. {\bf 234}, 5 (1971);
E. Fischbach and C. Talmadge. {\it The search for non-Newtonian gravity},
AIP Press, Springer-Verlag, New York (1998).
\bibitem{ICAN}
Gerard Mourou, Bill Brocklesby, Toshiki Tajima and Jens Limpert, 
Nature Photonics 7, 258–261 (2013).
\bibitem{PDGkine}
See Eq.(43.1) in
J. Beringer et al. (Particle Data Group), Phys. Rev. D86, 010001 (2012).
\bibitem{PDGstatistics}
See Eq.(36.56) in
J. Beringer et al. (Particle Data Group), Phys. Rev. D86, 010001 (2012).
\bibitem{ALPS}
K. Ehret et al. (ALPS Collab.), Phys. Lett. B689, 149 (2010).
\bibitem{Eto-wash}
E. G. Adelberger et al., Phys. Rev. Lett. 98, 131104 (2007);
D. J. Kapner et al., Phys. Rev. Lett. 98, 021101 (2007).
\bibitem{Stanford1}
J. Chiaverini et al., Phys. Rev. Lett. 90, 151101 (2003).
\bibitem{Stanford2}
S. J. Smullin et al., Phys. Rev. D 72, 122001 (2005) [Erratum-ibid. D 72, 129901 (2005)].
\bibitem{Lamoreaux}
S. K. Lamoreaux, Phys. Rev. Lett. 78, 5 (1997) [Erratum-ibid. 81:5475 (1998)].
\bibitem{QCDaxions}
See the parametrization, for exmple,
in the section {\it AXIONS AND OTHER SIMILAR PARTICLES} in
J. Beringer et al. (Particle Data Group), Phys. Rev. D86, 010001 (2012).
And also see
S. L. Cheng, C. Q. Geng, and W. T. Ni, Phys. Rev. D52, 3132 (1995)
on the possible range of the ratio on the electromagnetic to
color anomaly factors of the axial current associated with the axion, $E/N$.
\bibitem{KSVZ}
J. E. Kim, Phys. Rev. Lett.  43 , 103 (1979);
M. A. Shifman, A. I. Vainshtein and V. I. Zakharov, Nucl.
Phys. B 166 , 493 (1980).
\bibitem{Correction}
D. Bernard, arXiv:1311.5340 [physics.acc-ph].
\bibitem{BJ}
J. D. Bjorken and S. D. Drell, {\it Relativistic Quantum Mechanicsh},
McGraw-Hill, Inc. (1964);
See also Eq.(3.80) in W. Greiner and J. Reinhardt, 
{\it Quantum Electrodynamics Second Edition}, Springer (1994).
\bibitem{K-factor}
M. A. Furman, LBNL-53553, CBP Note-543.
\bibitem{Moeller}
C. M\"{o}ller, {\it General Properties of the Characteristic Matrix in the Theory of Elementary Particles. I} Munksgaard 1st edition (1946).
\bibitem{Chi3Gas}
J. F. Reintjes, {\it Nonlinear optical parametric processes in liquids and gases},
Academic Press, Orlando (1984).
\bibitem{AngularRep}
L. Novotny and B. Hecht {\it Principles of Nano-Optics second edition} 
Cambridge Univ. Press (2012);
A similar description can also be found in 
http://www.photonics.ethz.ch/fileadmin/user\_upload/optics/\\
Courses/EM\_FieldsAndWaves/AngularSpectrumRepresentation.pdf
\end{references}
\end{document}